\documentclass[twocolumn,secnumarabic,amssymb, nobibnotes, aps, prd]{revtex4}
\usepackage{graphicx}

\begin{document}
\title{Belief in thermodynamics has provoked false thermodynamics of superconductors}
\author{A.V. Nikulov}
\email[]{nikulov@iptm.ru}
\affiliation{Institute of Microelectronics Technology and High Purity Materials, Russian Academy of Sciences, 142432 Chernogolovka, Moscow District, RUSSIA.} 
\begin{abstract} Belief in thermodynamics has forced superconductivity experts to forget basics of thermodynamics due to a contradiction of superconductivity phenomena to laws of thermodynamics. Because of this belief no one drew reader's attention during many years that the conventional theory of superconductivity contradicts to the second law of thermodynamics. The common belief that the superconducting transition occurs when the free energy of the superconducting state becomes less than of the normal state has provoked a false claim that a power source of a solenoid creates the energy of magnetization rather than of magnetic field. The authors of only a few books on superconductivity, mostly future Nobel prize winners, did not follow this false claim. No one for many years has noticed that the equality of free energies at the superconducting transition in the critical magnetic field cannot be obtained without contradicting the second law of thermodynamics. The Meissner effect violates the second law of thermodynamics because of the negative surplus work performed in the closed Gorter cycle. The desire to avoid contradiction of superconductivity phenomena with the second law of thermodynamics provoked the false thermodynamics of superconductors, contradicting the law of conservation of energy.
\end{abstract}

\maketitle 


\section{Introduction}
The article of Jorge Hirsch \cite{Hirsch2024} attracts attention by its very title "Does the Meissner effect violate the second law of thermodynamics?". The fundamental and practical importance of this question is emphasized, for example, in a review article \cite{PhysRep1999}: {\it "The second law of thermodynamics is, without a doubt, one of the most perfect laws in physics. Any reproducible violation of it, however small, would bring the discoverer great riches as well as a trip to Stockholm. The world's energy problems would be solved at one stroke. It is not possible to find any other law (except, perhaps, for super selection rules such as charge conservation) for which a proposed violation would bring more skepticism than this one. Not even Maxwell's laws of electricity or Newton's law of gravitation are so sacrosanct, for each has measurable corrections coming from quantum effects or general relativity. The law has caught the attention of poets and philosophers and has been called the greatest scientific achievement of the nineteenth century"}. 

The possibility of solving the world's energy problems is explained by the fact that we must use fuel to generate energy from heat, only because of the Carnot principle, which we call the second law of thermodynamics since Clausius's time \cite{Smoluchowski}. Skepticism about the violation of this law is explained by the fact that Sadi Carnot based his principle on the centuries-old faith in the impossibility of a perpetual motion machine. Carnot wrote in his brilliant work of 1824 that "{\it it would be not only a perpetual motion, but also an unlimited creation of motive power without the cost of phlogiston or any other agents}" \cite{Carnot} if the efficiency $\eta = W/Q_{he}$ of a heat engine could exceed the maximum value 
$$\eta_{max} = (\frac{W}{Q_{he}})_{max} = 1 - \frac{T_{co}}{T_{he}}  \eqno{(1)}$$
The genius of Carnot is that using the notion about heat as a liquid, phlogiston, he determined the relationship between maximum efficiency of any heat engine and the impossibility of a perpetual motion machine, which is valid in various concepts of heat. According to the Carnot principle only a part of the heat obtained in the heater $W$ can be used for the work $W =  Q_{he} - Q_{co}$ while the other part $Q_{co} =  Q_{he} - W$ must transfer from the heater with a higher temperature $T_{he}$ to the cooler with a lower temperature $T_{co}$. Thus, we must use fuels to create a temperature difference $T_{he} - T_{co}$ between the heater $T_{he}$ and the cooler $T_{co}$ and we must increase the total heat in order to get useful energy from heat due to the faith in the impossibility of a perpetual motion machine. This faith cannot have a scientific basis because of the fundamental changes in our ideas about nature that have occurred over several centuries of its domination \cite{Entropy2022}. Therefore the Hirsch question \cite{Hirsch2024} is valid and should be discussed. 

But this question is not the main issue in this paper, since the contradiction of the Meissner effect to the second law of thermodynamics was proved several years ago \cite{Entropy2022}. More than twenty years ago, the violation of the second law of thermodynamics was proved in other macroscopic quantum phenomena observed in superconductors, see section 4.4 "Nikulov Inhomogeneous Loop" of the book \cite{book2005Capek} and the next section. The purpose of this article is to draw attention to the fact that belief in thermodynamics has provoked false thermodynamics of superconductors and a regression in the understanding of thermodynamics among superconductivity experts. One of the evidences of this regression is the amazing fact that all experts on superconductivity forgot for almost ninety years that the dissipation of kinetic energy of electric current into Joule heat is an irreversible process according to the second law of thermodynamics. 

The essence of this fact is considered in the next section. Section 3 draws attention to the fact that belief in thermodynamics has provoked contradictions with the law of conservation of energy in most books on superconductivity. The fourth section draws attention to the fact that the equality of free energies at the superconducting transition in the critical magnetic field written in most books on superconductivity was obtained at the cost of contradiction with the second law of thermodynamics. The fifth section draws attention to the contradictions between books on superconductivity that have been provoked by faith in thermodynamics and arbitrariness in the definition of free energy. The history of regress in the understanding of thermodynamics is considered in section six. The conclusion is made in Conclusion that the obvious falsity of the thermodynamics of  superconductors, which has been most clearly demonstrated in the contradictions between books on superconductivity, and the regression in the understanding of thermodynamics among experts on superconductivity are a consequence of the centuries-old belief in the impossibility of a perpetual motion machine.   

\section{Why the BCS theory of superconductivity contradicts to the second law of thermodynamics}
It should be emphasized that the question "Does the Meissner effect violate the second law of thermodynamics?" is not essential for J. Hirsch. J. Hirsch uses this question to prove the superiority of his alternative theory of hole superconductivity over the conventional BCS theory \cite{BCS1957}. He argues in the conclusion of his article \cite{Hirsch2024} "{\it that one of the following two alternatives has to be valid: (1) The Meissner effect violates the second law of thermodynamics, and is consistent with the BCS theory of superconductivity, as argued by Nikulov in \cite{Entropy2022}. (2) The Meissner effect is consistent with the second law of thermodynamics, establishes the invalidity of the BCS theory of superconductivity \cite{Hirsch2009} at least in its present form, and is consistent with the theory of hole superconductivity \cite{Hirsch2020book,HirschList,HirschOSF}}". These two alternatives mean that the BCS theory of superconductivity contradicts the second law of thermodynamics. 

The BCS theory was published in 1957 \cite{BCS1957} and in 1972 its authors, J. Bardeen, L.N. Cooper and J.R. Schrieffer (BCS), received the Nobel Prize in Physics. Many years have passed since then, and until recently, no one stated that this outstanding theory contradicts one of the basic laws of physics, which almost all scientists believe in. No one paid attention to this contradiction for a long time because all experts on superconductivity kind of forgot for many years that the dissipation of kinetic energy of electric current into Joule heat is an irreversible process according to the second law of thermodynamics. Hirsch recalled this basis of thermodynamics in three 2020 publications \cite{Hirsch2020Physica,Hirsch2020EPL,Hirsch2020ModPhys}. This reminder is of fundamental importance, because it is known since the discovery of the Meissner effect in 1933 \cite{Meissner1933} that superconducting currents that disappears after the transition of a superconductor to the normal state can reappear in a time-constant magnetic field upon return to the superconducting state.   

Some physicists understood ninety years ago that if superconducting currents disappear due to non-zero resistance in the normal state with the generation of Joule heat, the reappearance of these currents when the bulk superconductor returns to the superconducting state refutes the second law of thermodynamics. Therefore, they changed their opinion about the transition of a bulk superconductor to the normal state after the discovery of the Meissner effect in 1933 \cite{Meissner1933}. Before 1933 "{\it it was believed that the transition in a magnetic field is essentially irreversible, since the superconductor was considered as a perfect conductor (in the sense indicated in Chapter II), in which when the superconductivity is destroyed, the surface currents associated with the field are damped with the generation of Joule heat}" \cite{Shoenberg1938}. After 1933 all experts on superconductivity were sure that this transition is a first order phase transition.   

Any phase transition is by definition a reversible process in which the free energy cannot change and no irreversible processes such as the generation of Joule heat can be. The famous physicist W.H. Keesom wrote in 1934 that "{\it it is essential that the persistent currents have been annihilated before the material gets resistance, so that no Joule-heat is developed}" \cite{Keesom1934}. W.H. Keesom actually set the following task for the future theory of superconductivity: the theory should explain how the persistent currents can be annihilated before the material gets resistance. The authors of the conventional BCS theory of superconductivity \cite{BCS1957} not only did not solve, but did not even consider the Keesom task \cite{Keesom1934} although they, like all experts on superconductivity, were sure that superconducting transition is a phase transition. Hirsch was the first \cite{Hirsch2020Physica,Hirsch2020EPL,Hirsch2020ModPhys} to call attention to this internal contradiction in the BCS theory \cite{BCS1957}. The article \cite{Entropy2022} draws attention to the contradiction with the second law of thermodynamics because of the absence of a solution of the Keesom task in the BCS theory \cite{BCS1957}. 

J. Hirsch drew attention to the internal inconsistency \cite{Hirsch2020Physica,Hirsch2020EPL,Hirsch2020ModPhys} and contradiction with the second law of thermodynamics \cite{Hirsch2024} because of his belief that his alternative theory of superconductivity can solve the Keesom task, unlike the BCS theory \cite{BCS1957}. Because of his belief in the superiority of his theory, J. Hirsch ignores facts that call this superiority into question. The article \cite{PhysicaC2021} draws attention to experimental results \cite{LP1962,Science2007,Letter2007,toKulik2010,APL2016} which give experimental evidence that the persistent currents may not be annihilated not only before, but even after the material gets resistance. The Keesom task \cite{Keesom1934} lost its meaning due to experimental results \cite{PhysicaC2021}, obtained for the first time in 1962 \cite{LP1962}, after the death of W.H. Keesom. Numerous observations \cite{LP1962,Science2007,Letter2007,toKulik2010,APL2016} of the persistent currents $I_{p} \neq 0$ that are not annihilated in rings with nonzero resistance $R > 0$, despite the absence of the Faraday electromotive force $d\Phi /dt = 0$, is one of numerous experimental facts which refutes the superiority of the alternative theory of Hirsch \cite{Hirsch2020book,HirschList}. The other experimental fact is the quantization of magnetic flux.  

J. Hirsch states in a recent article \cite{Hirsch2025June} that "{\it The process by which the magnetic field is expelled killed in the transition from the normal to the superconducting state (Meissner effect) has not been described within the conventional theory}" and writes as his theory describes this process: "{\it The alternative theory of hole superconductivity \cite{Hirsch2020book,HirschList} predicts that metals expel negative charge from their interior to the surface when they become superconducting \cite{Hirsch2001}. In the presence of a magnetic field the outflowing charge acquires azimuthal velocity through the Lorentz force \cite{Hirsch2003} generating a magnetic field in direction opposite to the applied field, thus accounting for the Meissner effect. The theory also predicts that there is a back flow of negative charge during the transition to preserve near charge neutrality and to transfer azimuthal momentum of opposite sign to the body as a whole \cite{Hirsch2016PhS,Hirsch2017PRB}. The theory predicts that in the absence of radial charge flow magnetic fields would not be expelled. Instead, the conventional theory predicts no radial charge flow. We have discussed in previous work why radial charge flow is required to explain the Meissner effect \cite{Hirsch2016PhS,Hirsch2017PRB,Hirsch2010PhysC}}" \cite{Hirsch2025June}. 

It is surprising that J. Hirsch, when considering a superconducting cylinder with a hole in his article \cite{Hirsch2025June,Hirsch2025Aug}, forgot about the quantization of magnetic flux, which was observed in such cylinders. These observations, first in 1961 \cite{fluxquan1961a,fluxquan1961b}, together with the observation of quantum oscillations in magnetic field of the persistent current in the Little-Parkes effect  \cite{LP1962}, are the first direct experimental evidence that superconductivity is a macroscopic quantum phenomenon. Although few superconductivity experts doubted even before these observations that the Meissner effect is a macroscopic quantum phenomenon. Lev Landau was the first to explain in 1941 \cite{Landau41} the Meissner effect as a macroscopic quantum phenomenon using a wave function $\Psi _{GL} =|\Psi _{GL}|\exp{i\varphi }$ in which $|\Psi _{GL}|^{2} = n_{s}$ describes the density of superconducting mobile charge carriers and the phase gradient $\nabla \varphi = p/\hbar $ describes the canonical momentum $p = mv + qA$ of each carrier with a charge $q$ and a mass $m$. This wave function is the base of the Ginzburg-Landau theory \cite{GL1950}. 

L. Landau deduced the equation for the density of superconducting current $j = qn_{s}v = qn_{s}(\hbar \nabla \varphi - qA)/m$ from the equation for the velocity of a quantum particle $v = (\hbar \nabla \varphi - qA)/m$. This equation allowed not only to explain the Meissner effect but also to predict \cite{Entropy2022,PhysicaC2021} the observations of the flux quantization in superconducting cylinders with thick walls $w \gg \lambda _{L}$ \cite{fluxquan1961a,fluxquan1961b} and the persistent currents in a cylinder with thin walls $w \ll \lambda _{L}$ \cite{LP1962} or a ring with a small cross-section $s \ll \lambda _{L}^{2}$ \cite{PCJETP07}. $\lambda _{L} = (m/\mu _{0}q^{2}n_{s})^{0.5} \approx 50 \ nm = 5 \ 10^{-8} \ m$ is the London penetration depth. The relationship  $\oint_{l}dl j = (n\Phi_{0} - \Phi)/\mu _{0}\lambda _{L}^{2} $ between the current density $j$ along any closed path $l$, the magnetic flux $\oint_{l}dl A = \Phi $ inside this path and the quantum number $n$ is deduced from the Landau equation using the requirement $\oint_{l}dl \nabla \varphi = n2\pi $ that the wave function $\Psi _{GL} =|\Psi _{GL}|\exp{i\varphi }$ must be single-valued. 

This relationship predicted \cite{Entropy2022,PhysicaC2021} the persistent current $I_{p} = sj = (n\Phi_{0} - \Phi)s/\mu _{0}\lambda _{L}^{2}2\pi r = (n\Phi_{0} - \Phi)/L_{k}$ observed in a ring with a small cross-section $s \ll \lambda _{L}^{2}$ \cite{PCJETP07}. $L_{k} = (\lambda _{L}^{2}/s)\mu_{0}2\pi r = m2\pi r/sq^{2}n_{s} $ is the kinetic inductance of the ring with radius $r$, cross-section $s$ and density $n_{s}$ of superconducting particles with charge $q$. The quantization of magnetic flux $\Phi = n\Phi_{0}$ is observed \cite{fluxquan1961a,fluxquan1961b} since $j = 0$ inside thick walls $w \gg \lambda _{L}$. The quantum number $n$ can be non-zero if only a singularity of the wave function $\Psi _{GL} =|\Psi _{GL}|\exp{i\varphi }$, such as a hole in the superconducting cylinder \cite{fluxquan1961a,fluxquan1961b} or Abrikosov vortex \cite{Abrikosov} is inside the closed path $l$ \cite{Entropy2022,PhysicaC2021}. The flux quantization $\Phi = n\Phi_{0}$ is observed  \cite{fluxquan1961a,fluxquan1961b} at measurement of the superconducting cylinder with a hole and the Meissner effect \cite{Meissner1933} $\Phi = 0$ without a hole or other singularity of the wave function \cite{Entropy2022,PhysicaC2021}. Thus, according to the conventional theory of superconductivity \cite{GL1950} the Meissner effect $\Phi = 0$ is a special case of the flux quantization $\Phi = n\Phi_{0} $ when the quantum number $n = 0$. 

J. Hirsch is right \cite{Hirsch2025June} that the conventional theory has not described the process by which the magnetic field is expelled from a bulk superconductor \cite{Meissner1933}. But the Ginzburg-Landau theory \cite{GL1950} explains the appearance of the persistent currents in rings with the weak screening $s \ll \lambda _{L}^{2}$ \cite{PCJETP07,PRB2014}, cylinders with thick walls $w \gg \lambda _{L}$ \cite{fluxquan1961a,fluxquan1961b} and at the Meissner effect \cite{Meissner1933} in cylinders without a hole as a consequence of quantization: the persistent currents must appear when superconducting state with zero current is forbidden \cite{PhysicaC2021}. According to the Hirsch theory of hole superconductivity \cite{Hirsch2020book,HirschList} the persistent currents appear under influence of the Lorentz force \cite{Hirsch2003} rather than because of quantization. But the Lorentz force cannot in principle explain the appearance of the persistent currents not only in a cylinder with thin walls $w \ll \lambda _{L}$ \cite{LP1962} or a ring with a small cross-section $s \ll \lambda _{L}^{2}$ \cite{PCJETP07}, where radial charge flow cannot be possible geometrically, but also in cylinders with thick walls $w \gg \lambda _{L}$ \cite{fluxquan1961a,fluxquan1961b}, since the persistent currents $I_{p} = sqn_{s}v = (n\Phi_{0} - \Phi)/L_{k}$ \cite{PCJETP07} and flux quantization $\Phi = n\Phi_{0}$  \cite{fluxquan1961a,fluxquan1961b} are quantum phenomena. 

J. Hirsch may insist on the superiority of his theory only because he has forgotten that superconductivity is a macroscopic quantum phenomenon. He is right that the conventional theory \cite{BCS1957,GL1950} cannot removed the contradiction of the Meissner effect with Faraday's law and the law of angular momentum conservation. This contradiction is observed also at the appearance of a macroscopic persistent current $I_{p} = sqn_{s}v = qN_{s}v/2\pi r = (n\Phi_{0} - \Phi)/L_{k}$ \cite{PCJETP07} in a ring at its transition in superconducting state \cite{PRB2014} with a macroscopic number of superconducting pairs $N_{s} = s2\pi r n_{s}$ when a magnetic flux inside the ring is not divisible $\Phi = \pi r^{2}\mu_{0}H \neq n\Phi_{0}$ by the flux quantum $\Phi _{0} = 2\pi \hbar /q$. Comparison of the persistent currents of electrons observed in normal metal rings \cite{PCScien09,PCPRL09} and superconducting pairs \cite{PCJETP07} allowed to conclude that this contradiction is a consequence of violation of the correspondence principle in macroscopic quantum phenomena \cite{ChJoPh2024}. The period oscillations in magnetic field $\mu_{0}H_{0} = B_{0} = \Phi _{0}/\pi r^{2}$ of the both persistent currents equals the flux quantum $\Phi _{0} = 2\pi \hbar /q$ inside the ring with a radius $r$. The value of the flux quantum $\Phi _{0} = 2\pi \hbar /2e \approx 20.7 \ Oe \ \mu m^{2}$ corresponds the charge of Cooper pair $q = 2e$ for superconducting rings \cite{PCJETP07} and $\Phi _{0} = 2\pi \hbar /e \approx 41.4 \ Oe \ \mu m^{2}$ corresponds the charge of electron $q = e$ for normal metal rings \cite{PCScien09,PCPRL09}.          

The quantum periodicity is experimental evidence that the persistent currents are observed due to the quantization of the angular momentum $rp = rmv + rqA = rmv + q\Phi /2\pi = rmv + \hbar \Phi /\Phi _{0} = n\hbar $, according to which the velocity $v = \hbar (n - \Phi /\Phi _{0})/rm$ cannot be equal zero when the magnetic flux inside the ring is not divisible $\Phi \neq n\Phi _{0}$ to the flux quantum \cite{PhysicaC2021}. Since the average velocity is zero in the absence of quantization, it should change by $\Delta v = \hbar (n - \Phi /\Phi _{0})/rm$ when quantization appears. The persistent current observed in normal metal rings \cite{PCScien09,PCPRL09} does not contradict to the law of conservation since the change of angular momentum of electron $\Delta p_{r} = rmv = \hbar (n - \Phi /\Phi _{0}) < \hbar$ does not exceed Planck's constant in accordance with the correspondence principle \cite{ChJoPh2024}. The angular momentum of Cooper pairs, in contrast to the electron, does not change individually, but all at once, according to both the conventional theory of superconductivity \cite{BCS1957,GL1950} and all experimental results. The appearance of even the relatively small persistent current $I_{p} \approx 10 \ \mu A = 10^{-5} \ A$, observed in the rings with a radius $r \approx 2 \ \mu m$ \cite{PCJETP07}, corresponds to the change of the angular momentum by a macroscopic amount $\Delta P_{r} = N_{s}rmv = N_{s}\hbar (n - \Phi /\Phi _{0}) = I_{p}2\pi r^{2}(m/e) \approx  2 \ 10^{10} \ \hbar $ because of a huge number of Cooper pairs $N_{s} = Vn_{s} = s2\pi rn_{s} > 10^{10} $ in a real ring with a macroscopic volume $V = s2\pi r \approx 0.2 \ \mu m^{3}$ \cite{PCJETP07}.   

No one has noticed that macroscopic quantum phenomena cannot be observed in accordance with the correspondence principle. Also, no one noticed for many years that the change in angular momentum by a macroscopic amount in the absence of a force observed in these phenomena contradicts the law of conservation. J. Hirsch noticed the contradiction, but he did not notice that the problem with the conservation law is a consequence of a violation of the correspondence principle because of his belief in the superiority of his theory of superconductivity. The statement of the conventional theory of superconductivity \cite{BCS1957,GL1950} about the change of angular momentum by a macroscopic amount $\Delta P_{r} = N_{s}rmv \gg \hbar $ due to quantization in the absence of force contradicts the conservation law. But the theory is not false because of this contradiction, since the contradiction with the conservation law is observed experimentally.

The conventional theory of superconductivity \cite{BCS1957,GL1950} is not false also because of its contradiction with the second law of thermodynamics \cite{Entropy2022}. This contradiction is also a consequence of the violation of the correspondence principle \cite{ChJoPh2024}. A macroscopic kinetic energy $E_{k} = I_{p,A}\Phi_{0}(n- \Phi /\Phi_{0})^{2}$ emerges because of the quantization and contrary to this principle when a ring becomes superconducting \cite{Entropy2022,PhysicaC2021}. This kinetic energy can exceed strongly the fluctuation energy $k_{B}T$: $E_{k} \approx  1000k_{B}T$ \cite{NanoLet2017} at the relatively small persistent current $I_{p} = I_{p,A}2(n- \Phi /\Phi_{0}) \approx 10 \ \mu A = 10^{-5} \ A$ \cite{PCJETP07}. J. Hirsch wrote correctly that according to the conventional theory \cite{BCS1957,GL1950} this energy dissipates into Joule heat as a result of electron scattering after returning to the normal state: "{\it As the system becomes normal, Cooper pairs unbind and become normal quasiparticles, and the supercurrent stops. Within the conventional theory this process has been discussed by Eilenberger \cite{Eilenberger1970} using the time-dependent Ginzburg-Landau (TDGL) formalism. A term in the current density describes the current carried by normal electrons stemming from the momentum transferred to the normal electron fluid when the superfluid electron density decreases. Eilenberger states that "this momentum then decays with the transport relaxation time $\tau $"}" \cite{Hirsch2017PRB}. 

The reappearance of the macroscopic kinetic energy $E_{k} \gg k_{B}T$ is a process revers to its dissipation into Joule heat, which must be irreversible according to the second law of thermodynamics. Only a solution of the Keesom task \cite{Keesom1934} could save this fundamental law from experimental refutation. Not only conventional theory \cite{BCS1957,GL1950}, but also Hirsch's theory cannot solve this task. Moreover, the solution to this task has lost its meaning after the numerous observation of the persistent currents $I_{p} \neq 0$ at a nonzero resistance $R > 0$  \cite{LP1962,Science2007,Letter2007,toKulik2010,APL2016}. This paradoxical phenomenon is observed only a narrow temperature region near superconducting transition $T \approx  T_{c}$ due to thermal fluctuations \cite{Skocpol1975}: $I_{p} \neq 0$ but $R = 0$ at $T < T_{c}$ and  $R > 0$ but $I_{p} = 0$ at $T > T_{c}$ if the conventional theory of superconductivity \cite{BCS1957,GL1950} does not taken into account thermal fluctuations.

Studies of thermal fluctuations, in particular Brownian motion, fundamentally changed in the beginning of the 20th century views on the essence of the second law of thermodynamics \cite{Entropy2022}. M. Smoluchowski, who, along with Einstein, made important contributions to the theories of Brownian motion, wrote in his article published in 1914: "{\it In the phenomena of fluctuation experimentally observed in recent years, it seems extremely strange for supporters of classical thermodynamics that he sees with his own eyes the reverse course of processes that are generally regarded as irreversible. Because according to the classical theory, the second law of thermodynamics should disappear if at least one process, regarded as irreversible, admits reversibility}" \cite{Smoluchowski}. The phenomena of fluctuation seems extremely strange for supporters of classical thermodynamics since they  denied not only the perpetual thermal motion of atoms, but the very existence of atoms, because of their belief in the second law of thermodynamics \cite{Smoluchowski}. 

Their view prevailed in the end of the 19th century and was abandoned due to the study of fluctuations. Smoluchowski wrote in 1914: "{\it Thus, the issue is considered very different today than it was twenty years ago. Atomistics is recognized as the basis of modern physics in general; the second law of thermodynamics has once and for all lost its significance as an unshakable dogma, as one of the basic principles of physics}" \cite{Smoluchowski}. But the dogma has changed rather than lost its significance because of common belief in the impossibility of a perpetual motion machine. Smolukhovsky argued: "{\it On the contrary, from the point of view of molecular statistics, the position of thermodynamics about the impossibility of a perpetual motion machine of the second kind is correct, if this expression is given a more precise meaning, namely: 'an automatic machine continuously making work by consuming the heat of another body with a lower temperature'}" \cite{Smoluchowski}.   

Smoluchowski stated that "{\it the widespread opinion that molecular fluctuations could be used directly to construct a simple perpetuum mobile is completely wrong}" \cite{Smoluchowski} and tried to prove his statement by the impossibility of creating a directed thermal motion. He proved that a directed thermal motion cannot be created with the help of the mechanical machine since all parts of this machine move chaotically due to thermal fluctuations \cite{Smoluchowski}. Richard Feynman repeated this proof of Smoluchowski in Chapter 46 "Ratchet and pawl" of his lectures on physics \cite{Feynman1963}. But if a directed thermal motion cannot be created with Feynman's ratchet \cite{Feynman1963}, it cannot prove the impossibility of such a motion in all cases. The great scientists Max Planck, unlike Smoluchowski, Feynman and most physicists, understood that the impossibility of a directed thermal motion may be only an assumption, which cannot be proved. He questioned the Boltzmann H - theorem because of the groundlessness of this assumption: "{\it Boltzmann omitted in his deduction every mention of the indispensable presupposition of the validity of his theorem namely, the assumption of molecular disorder. He must have simply taken it for granted}" \cite{Planck}.

The article \cite{Entropy2022} draws attention to the fact that Planck was not quite right. Boltzmann understood the need of the assumption of molecular disorder for the validity of his theorem. But most modern theorists not only have simply taken this assumption for granted, but also do not understand its need for the validity of the Boltzmann H - theorem, and as consequence of the second law of thermodynamics \cite{Entropy2022}. The regression in the understanding of the conditions necessary for the validity of the second law of thermodynamics is evidenced by the following fact. The article in which the persistent current $I_{p} \neq 0$ observed at non-zero resistance $R > 0$ \cite{LP1962} was explained as directed Brownian motion was published \cite{PRB2001}. But the Comment to this article \cite{PRB2001} and the Reply \cite{Reply2003}, in which it was reminded that the observation of directed Brownian motion is a challenge to the second law of thermodynamics, were banned from publication. It was shown back in 2002 that this directed Brownian motion has nothing to do with either Feynman's ratchet or Maxwell's demon \cite{AIP2002my}. The violation of the assumption of molecular disorder is observed in the publications \cite{LP1962,Science2007,Letter2007,toKulik2010,APL2016,PCScien09,PCPRL09} because of the discreteness of permitted state spectrum of electrons in normal metal rings with enough small radius \cite{PCScien09,PCPRL09} and of Cooper pairs in superconducting rings \cite{LP1962,Science2007,Letter2007,toKulik2010,APL2016}. The discreteness \cite{PLA2012T} is not subject to chaotic thermal movement as opposed to Feynman's ratchet \cite{FQMT2004}.            

Due to of this discreteness the power $W_{pc} = \overline{RI_{p}^{2}}$ observed in the phenomena of the persistent current is not zero at the zero frequency $\omega = 0$ \cite{AIP2002my} in contrast \cite{Entropy2022,PhysicaC2021,FQMT2004} to such well known chaotic Brownian motion \cite{Feynman1963} as the Nyquist \cite{Nyquist} (or Johnson \cite{Johnson}) noise. It is known since 1928 \cite{Nyquist,Johnson} that any element of an electric circuit can be a power source $W_{Nyq} = \overline{V^{2}}/R$ at a finite temperature $T$ due to the Nyquist noise. But this power cannot be used for an useful work under equilibrium conditions since the power of each element is distributed among the same frequency $\omega$ spectrum \cite{FQMT2004}: $W_{Nyq} = 4k_{B}T\Delta \omega $ from $\omega = 0$ up to the quantum limit $\omega < k_{B}T/\hbar$ \cite{Feynman1963}. We can not say what element is the power source and which element is the load. The observations \cite{Letter2007,APL2016,LP1962,Science2007,toKulik2010} of the persistent current $I_{p} \neq 0$ which is not damped in spite of a non-zero resistance $R > 0$ contradicts to the second law of thermodynamics since this paradoxical phenomenon give evidence of a dc power source $W_{pc} = \overline{RI_{p}^{2}}$, which can be used for a useful work at thermodynamic equilibrium \cite{Entropy2022,PhysicaC2021,AIP2002my}. 

The conventional theory of superconductivity \cite{GL1950} not only can explain \cite{PRB2001} the observation of the persistent power $W_{pc} = \overline{RI_{p}^{2}}$  \cite{Letter2007,APL2016,LP1962,Science2007,toKulik2010} but also can predict a possibility to observe the persistent voltage $V_{p} \propto I_{p}$ and the persistent power source $W_{pv} = \overline{V_{p}I_{p}} = \overline{V_{p}^{2}/R}$ on segments of asymmetric superconducting rings. The possibility of such prediction was shown in 1998 \cite{LTP1998} and was repeated in 2024 \cite{Berger2024}. A voltage with a dc component proportional to the persistent current $V_{dc} \propto I_{p}$ can be observed on segment of a superconducting ring when this segment is switched between superconducting and normal state \cite{PhysicaC2021} according to \cite{LTP1998}. This prediction was confirmed experimentally \cite{Letter2007,APL2016,PCJETP07,NANO2002,Letter2003,PLA2012Ex,PLA2017} in spite of the fact that the observation of the persistent current flowing against the action of a constant electric field $E_{dc} = -\nabla V_{dc}$ in one of the ring segments \cite{Physica2019} contradicts the law of conservation of momentum.   

The contradiction with Faraday's law and the law of angular momentum conservation in the observations \cite{Letter2007,APL2016,PCJETP07,NANO2002,Letter2003,PLA2012Ex,PLA2017,Physica2019} of the dc voltage  $V_{dc} \propto I_{p}$ is no less, and even more obvious, that in the case of the Meissner effect. The conventional theory of superconductivity \cite{GL1950} can explain even these paradoxical observation as a consequence of quantization. The potential difference $V(t) = R_ {B}I(t) = R_ {B}I_{p}\exp -t /\tau_{RL} $ should appear on the half $B$ of a superconducting ring with the persistent current $I_{p}$ when this half is switched in the normal state with a resistance $R_ {B}$ \cite{PhysicaC2021}. This voltage and the circular current $I(t) = I_{p}\exp -t /\tau_{RL}$ should decay, during a shot relaxation time $\tau_{RL} = L /R_ {B}$ determined by the resistance $R_ {B}$ and the inductance $L$ of the ring. The circular current should decay because of the dissipation force acting on electrons in the half $B$ and because of the electric field $E(t) = \nabla V(t)$ acting on Cooper pairs in the half $A$ remaining superconducting \cite{PhysicaC2021}. The persistent current $I_{p} = s2en_{s}v = s2en_{s}\hbar (n - \Phi /\Phi _{0})/rm $ should reappear because of quantization when the half $B$ returns to the superconducting state if the velocity of Cooper pairs $v = \hbar (n - \Phi /\Phi _{0})/rm$ cannot be zero at $\Phi \neq n\Phi _{0}$. The voltage average $V _{dc} =  \int _{0}^{\Theta }dt V(t)/\Theta = R _{B}\overline{I_{p}}f _{sw} \int _{0}^{t _{n}}dt \exp {-t/\tau_{RL}}$ during a long time $\Theta \gg 1/f _{sw}$ should be observed when the half $B$ is switched between superconducting and normal states with a frequency $f _{sw} = N _{sw}/\Theta $. Here $\overline{I_{p}}= \sum^{i=N _{sw}}_{i=1} I_{p,i}/N _{sw}$  is the average value of the persistent current after $N _{sw}$ returning of the half $B$ in the superconducting state during the time $\Theta \gg 1/f _{sw}$;  $t _{n}$ is the time during which the half $B$ is in the normal state; the integral $\int_{0}^{t_{n}} dt\exp {-t /\tau_{RL}} \approx  \tau_{RL} = L /R_{B}$ when $t_{n} \gg \tau_{RL}$ and $\int_{0}^{t_{n}} dt\exp {-t /\tau_{RL}} \approx t_{n}$ when $t_{n} \ll \tau_{RL}$. 

The dc voltage $V_{dc}(\Phi /\Phi _{0}) \propto I_{p}(\Phi /\Phi _{0})$ oscillating with the period corresponds the flux quantum $\Phi _{0} = 2\pi \hbar /2e \approx 20.7 \ Oe \ \mu m^{2}$ was observed as far back as 1967 \cite{Physica1967}. The authors \cite{Physica1967} were concerned only with the question about a source of the power $V _{dc}\overline{I_{p}} = V _{dc}^{2}/R$ that they observed. They did not notice that the dc voltage $V _{dc}(\Phi ) \propto \overline{I_{p}}(\Phi )$ is observed in a magnetic field constant in time $d\Phi /dt = \pi r^{2}\mu _{0} dH/dt = 0$, i.e. at the zero Faraday electric field $E_{F} = -dA/dt = -(1/2\pi r)d\Phi /dt = 0$. They assumed that the source of this power is the emitted electromagnetic radiation of broadcasting stations and have confirmed that the visible dc voltage $V _{dc}(\Phi ) \propto \overline{I_{p}}(\Phi )$ disappears when all parts of the equipment are shielded more carefully \cite{Physica1967}. The authors \cite{Physica1967} did not take into account that no electromagnetic radiation can explain how the persistent current can flow against the action of the constant electric field $E = E_{dc} + E_{F} = -\nabla V_{dc} - dA/dt = -\nabla V_{dc}$. 

The electromagnetic radiation like any non-equilibrium noise and equilibrium noise can only switch a ring or any loop with the persistent current between superconducting and normal state. Such switching is possible when the power of the noise can induce in the ring a current exceeding the critical current $I_{c}$. Measurements \cite{PCJETP07,Letter2003} have shown that the quantum oscillations $V_{dc}(\Phi /\Phi _{0}) \propto I_{p}(\Phi /\Phi _{0})$ are observed when the amplitude $I_{A,noise}$ of the noise current or the ac current exceeds the value of the critical current at a given temperature $I_{A,noise} > I_{c}(T)$. According to the conventional theory of superconductivity \cite{BCS1957,GL1950} and experimental results \cite{PCJETP07} the critical current $I_{c}(T) = I_{c}(T = 0)(1 - T/T_{c})^{3/2}$ increases with the temperature decrease, like the persistent current $I_{p}(T) = I_{p}(T = 0)(1 - T/T_{c})$. Therefore the quantum oscillations $V_{dc}(\Phi /\Phi _{0}) \propto I_{p}(\Phi /\Phi _{0})$ are observed only in a narrow temperature region at a small amplitude $I_{A,noise}$ of the noise current \cite{Letter2007,APL2016,PLA2012Ex}: at the lower temperature the dc voltage $V_{dc}(\Phi /\Phi _{0})$ is not observed because $I_{A,noise} < I_{c}(T = 0)(1 - T/T_{c})^{3/2}$ and at the high temperature because of its decrease with the decrease of the persistent current $I_{p}(T) = I_{p}(T = 0)(1 - T/T_{c})$. 

Because of the common belief in the second law of thermodynamics the authors \cite{Physica1967} falsely decided that the source of the observed dc power $W_{dc} = V _{dc}\overline{I_{p}} = V _{dc}^{2}/R$ is non-equilibrium noise and that this power cannot be observed at thermodynamic equilibrium. The maximum voltage $V _{dc,max}$ and the maximum power $W_{dc,max} = V _{dc,max}^{2}/R$ indeed decrease with the decrease of the amplitude $I_{A,noise}$ of the noise current. The noise current with the amplitude no less than $I_{A,noise} \approx  30 \ \mu A = 3 \ 10^{-5} \ A$ induced the quantum oscillations $V_{dc}(\Phi /\Phi _{0}) \propto I_{p}(\Phi /\Phi _{0})$ with maximum amplitude $V _{dc,max} \approx 15 \ \mu V = 1.5 \ 10^{-5} \ V$ and the dc power $W_{dc,max} = V _{dc,max}^{2}/R \approx 10^{-10} \ W$ in the 1967 work \cite{Physica1967}. The weaker noise $I_{A,noise} \geq 1 \ \mu A = 10^{-6} \ A$ induced on a single aluminum asymmetric ring the voltage $V _{dc,max} \approx 1.2 \ \mu V = 1.2 \ 10^{-6} \ V$ and the power $W_{dc,max} \approx 2 \ 10^{-13} \ W$ \cite{NANO2002}. The oscillation $V_{dc}(\Phi /\Phi _{0}) \propto I_{p}(\Phi /\Phi _{0})$ ceased to be visible on one ring when the noise amplitude was reduced with the help of screening and filtration. But this does not mean that the DC power $W_{dc} =  V _{dc}^{2}/R$ has completely disappeared \cite{Physica2019}. The oscillations with the amplitude $V _{dc,max} \approx 0.6 \ \mu V = 0.6 \ 10^{-6} \ V$ were observed on the system of 110 asymmetric aluminum rings \cite{Letter2007} due to that the DC voltage and the DC power are added up in the system of series-connected sources \cite{Physica2019}. The system of 667 asymmetric aluminum rings was used in \cite{APL2016} in order to detect the noise current with the amplitude $I_{A,noise} \geq 50 \ nA = 5 \ 10^{-8} \ A$. This noise induced the maximum voltage $V _{dc,max} \approx 1.5 \ nV = 1.5 \ 10^{-9} \ V$ and the maximum power $W_{dc,max} \approx 2 \ 10^{-17} \ W$ on one ring, if all 667 rings contribute to the voltage $V _{dc,max} \approx 1 \ \mu V = 10^{-6} \ V$ observed on the system.  

The decrease in the DC power $W_{dc} = V _{dc}^{2}/R$ with the decrease in the amplitude $I_{A,noise}$ of the noise current does not mean that noise is the source of energy for the observed DC power, contrary to the false conclusion made by the authors \cite {Physica1967}. The source of this energy is the kinetic energy of the persistent current. Each time the ring returns to the superconducting state, the persistent current $I_{p} = qsn_{s}v_{n} =  I_{p,A}(n- \Phi /\Phi_{0}) $ and the macroscopic kinetic energy $E_{k} = N_{s}mv_{n}^{2}/2 = L_{k}I_{p}^{2}/2 =  I_{p,A}\Phi_{0}(n- \Phi /\Phi_{0})^{2}$ reappear in it due to quantization. $I_{p,A} = \Phi_{0}/2L_{k}$ is the amplitude of the oscillations of the persistent current. The voltage $V_{dc} \propto \overline{I_{p}}$ is observed \cite{Letter2007,APL2016,PCJETP07,NANO2002,Letter2003,PLA2012Ex,PLA2017,Physica2019} because each time after the transition to the normal state, the current $I(t) = I_{p}\exp -t /\tau_{RL}$ induce the voltage $V(t) = 0.5(R_{B} - R_{A})I(t)$ on the halves of the ring with different resistance $R_{B} > R_{A}$ during the relaxation time $\tau_{RL}$.

The asymmetric rings can be switched by both non-equilibrium and equilibrium noise (thermal fluctuations). But thermal fluctuations can switch the ring to the normal state only near the superconducting transition $T \approx T_{c}$, where the persistent current $I_{p}(T) \approx  I_{p}(T = 0)(1 - T/T_{c})$ is very small. Therefore, a system with a sufficiently large number of asymmetric rings is needed to observe the power $W_{dc} = V _{dc}^{2}/R$ at thermodynamic equilibrium. The power $W_{dc} \approx  10^{-14} \ W$ observed on the system of 1080 rings \cite{PLA2012Ex} is close in magnitude to the power which should be observed at thermodynamic equilibrium \cite{arXiv2024}, according to calculations \cite{Berger2024} made on the basis of the Ginzburg-Landau theory \cite{GL1950}. But it should be emphasized that the observation of the DC power $W_{dc} = V _{dc}\overline{I_{p}} = V _{dc}^{2}/R$ refutes the second law of thermodynamics no matter what the reason for switching of the asymmetric rings between superconducting and normal states. The second law of thermodynamics is violated every time when heat is converted into the kinetic energy $E_{k} = I_{p,A}\Phi_{0}(n- \Phi /\Phi_{0})^{2}$ of the persistent current $I_{p}$.

The observed contradictions with the second law of thermodynamics and with the law of conservation of momentum are related to each other, predicted by the conventional theory of superconductivity \cite{BCS1957,GL1950} and are a consequence of quantization and violation of the correspondence principle \cite{ChJoPh2024}. Hirsch proposes to remove these contradictions with the help of the magnetic Lorentz force. He writes: "{\it Nikulov also argues that observation of dc voltages in asymmetric superconducting rings \cite{Physica2019} is another example of violation of the second law of thermodynamics by superconductors. Whether the theory of hole superconductivity can explain such observations consistent with thermodynamics is an open question}" \cite{Hirsch2024}. This question cannot be consider as open. The Hirsch theory, in contrast to the conventional theory of \cite{BCS1957,GL1950}, cannot explain in principle how the half of asymmetric ring can be the DC power source $W_{dc} = V _{dc}^{2}/R$ \cite{Letter2007,APL2016,PCJETP07,NANO2002,Letter2003,PLA2012Ex,PLA2017} in which current $\overline{I_{p}}$ flows against electric field $E_{dc} = -\nabla V_{dc}$ \cite{Physica2019}.      

\section{False claim of Gorter and Casimir.}
Almost no one pays attention to the experimental evidences \cite{LP1962,Science2007,Letter2007,toKulik2010,APL2016,PLA2012Ex} of violations of the second law of thermodynamics predicted \cite{PRB2001,LTP1998,Berger2024} by the conventional theory of superconductivity \cite{BCS1957,GL1950}, since this law has always been a matter of faith rather than understanding. Most scientists in the late 19th and even early 20th century negatively related to the Maxwell-Boltzmann statistical theory and many scientists, supporters of the thermodynamic-energy worldview denied even the existence of atoms and their perpetual thermal motion because of this faith. Smoluchowski wrote in his article "Limits of Validity of the Second Law of Thermodynamics" published in 1914: "{\it I begin my presentation of the above topic with a brief historical overview. Anyone who has been involved in the struggle between thermodynamic-energy and atomistic-kinetic worldviews for the past forty years knows why I do this. It is no longer easy for us to imagine the way of thinking that prevailed at the end of the last century. After all, at that time, scientists in Germany and France were convinced that the kinetic theory of atoms had already played a role. The principle of Carnot, intuitively understood by him, we call since the time of Clausius the second law of thermodynamics. Because of the confidence in the great achievements of thermodynamics, this principle has been elevated to the rank of the absolute, exact dogma without exclusion. And since at that time molecular kinetics in the interpretation of this principle faced certain difficulties associated with the irreversibility of processes, it, together with atomistics, was immediately condemned as untenable. Although Boltzmann tried to prove that if there are contradictions, they still cannot practically become tangible}" \cite{Smoluchowski}.

Carnot has understood his principle not entirely intuitively, but based it on the universal belief in the impossibility of a perpetual motion machine. That is why the second law of thermodynamics has been a matter of faith rather than understanding, for two centuries now. Some supporters of the thermodynamic-energy worldview admitted the perpetual motion of atoms because of the undeniable experimental evidence. The faith of 20th century physicists turned out to be more blind. Because of the blind faith in thermodynamics, many physicists forgot the basics of thermodynamics and created a false thermodynamics of superconductors, which everyone believed in for ninety years. The falsity of this thermodynamics is especially evident because of the contradictions with the law of conservation of energy in most books on superconductivity and the contradictions between the books. To understand how these contradictions could appear, it is necessary to recall the essence of the second law of thermodynamics. 

In the time of Carnot heat was considered a liquid, phlogiston. When heat began to be considered a type of energy in Clausius's time, it became clear that Carnot had postulated irreversibility in physics. The possibility of a perpetual motion machine would be inevitable according to the law of conservation of energy if all processes in physics were reversible. Rudolf Clausius formulated in the 1850s the irreversibility of the transformation of energy into heat, postulated by Carnot, as the second law of thermodynamics or the law of entropy increase through the concept of entropy, which he defined as the ratio $S = Q/T$ of heat $Q$ to temperature $T$. According to Clausius, the transformation of any other energy $E$ into heat $Q$ is a irreversible process since the entropy of only heat is not zero and therefore the total energy increases on $\Delta S = \Delta Q/T = E/T$ at this transformation. 

For example, when the kinetic energy $E_{k}$ of the persistent current dissipates into Joule heat the total entropy increases by $\Delta S = E_{k}/T$. The reverse process contradicts the second law of thermodynamics, since the total entropy in this process decreases by the same amount $\Delta S = -E_{k}/T$. Conversion of heat $Q_{he} - Q_{co}$ into energy $W$ in a heat engine does not contradict the second law of thermodynamics, since the decrease in entropy $\Delta S_{de} = -W/T_{he} =  -(Q_{he} - Q_{co})/T_{he}$ during the conversion of a part $Q_{he} - Q_{co}$ of the heater's heat $Q_{he}$ into energy $W = Q_{he} - Q_{co}$ is compensated by increase in the entropy $\Delta S_{in} = Q_{co}/T_{co} - Q_{co}/T_{he} = (Q_{he} - W)/T_{co} - (Q_{he} - W)/T_{he}$ due to the transfer of the other part of the heat $Q_{co} =  Q_{he} - W$ from the heater with a higher temperature $T_{he}$ to the cooler with a lower temperature $T_{co}$. The total entropy may not decrease $\Delta S_{tot} = \Delta S_{de} + \Delta S_{in} =  -  W/T_{he} + (Q_{he} - W)/T_{co} - (Q_{he} - W)/T_{he} = Q_{he}/T_{co} - Q_{he}/T_{he} - W/T_{co} \geq 0$ if $1 - T_{co}/T_{he} - W/Q_{he} \geq 0$ i.e. if the efficiency of the heat engine does exceed the maximum value (1) $\eta = W/Q_{he} \leq 1 - T_{co}/T_{he}$.
    
Free energy $F = U - ST$ is defined as a quantity that cannot increase in a closed thermodynamic system according to the second law of thermodynamics. Therefore free energy should not change at any reversible process, in particular during a phase transition. For example, the total energy $U = F + ST$ is not changed $\Delta U = 0$ and the free energy decrease on $\Delta F = \Delta U - \Delta ST = -E_{k}$ when the kinetic energy $E_{k}$ of the persistent current dissipates into Joule heat. The decrease in free energy and the impossibility of its increase in a closed thermodynamic system means that any system tends to thermodynamic equilibrium in which free energy is minimal. For more than a hundred years, no one doubted that these basics of thermodynamics could be applied to the phenomena of superconductivity. D. Shoenberg wrote in the book \cite{Shoenberg1938} published in 1938: "{\it The idea of applying thermodynamics to the transition from the superconducting to the normal state was first expressed by Keesom (1924 Rapp. 4-A Congr. Phys. Solvay, 288) and then by Rutgers \cite{Rutgers1933} and developed in detail by Gorter \cite{Gorter1934}}".  

The doubts about the reversibility of the superconducting transition that existed before the discovery of the Meissner effect disappeared after 1933: "{\it This supposed irreversibility made Gorter's interpretation questionable, and it seemed surprising that his results were in good agreement with experience. The reason for this agreement became clear later, after the discovery of the Meissner effect, which showed that the disappearance of surface currents in a pure metal is not actually associated with any irreversible energy transformations (somewhat analogous to the disappearance of surface currents in a ferromagnet at the Curie point), so Gorter's basic assumption of reversibility turns out to be correct}" \cite{Shoenberg1938}. According to the common belief, which no one has questioned for ninety years, the transition to the superconducting state occurs when the free energy of the superconducting state becomes less than the normal state. 

This belief implies that the following equations should be correct: $F_{s0} > F_{n0}$ at $T > T_{c}$ and $F_{s0} < F_{n0}$ at $T < T_{c}$ without magnetic field $H = 0$, and $F_{sH} < F_{nH}$ at $H < H_{c}(T)$ and $F_{sH} > F_{nH}$ at $H > H_{c}(T)$ in a magnetic field. $F_{s0}$ and $F_{n0}$ are the free energy of the superconducting and the normal states without magnetic field $H = 0$;  $T_{c}$ is the critical temperature at which the superconducting transition is observed at $H = 0$; $F_{sH}$ and $F_{nH}$ are the free energy of the superconducting and normal states in a magnetic field $H$ at $T < T_{c}$; $H_{c}(T)$ is the critical magnetic field at $T < T_{c}$. According to the common belief after 1933 the transition without magnetic field at $T = T_{c}$ is a second order phase transition while the transition of a bulk superconductor at the critical magnetic field $H = H_{c}(T)$ is a first order phase transition. This transition is considered a reversible process in spite the jump $S_{n} - S_{s} \neq 0$ in the entropy, equal to the first derivative of free energy with respect to temperature $S = - dF/dT$, since the change in heat $\Delta Q = L $, called the latent heat $L = T(S_{n} - S_{s})$, is compensated by the change in total energy $\Delta U = L $ as a result of internal processes in the thermodynamic system: $\Delta F = \Delta U - L = \Delta U - T(S_{n} - S_{s}) = 0$ and therefore $F_{sH} = F_{nH}$. 

A positive work can increase and a negative work can decrease the total energy $U = F + ST$. Free energy $F = U - ST$ may be defined also as the energy capable of performing work without processes revers to irreversible process. A positive work can increase both free energy and heat. But a negative work can use only free energy but not heat. A power source of a solenoid does positive work to create a magnetic field $H$ in a volume $V$,  
$$W = \int _{t}dt IV = V\int _{t}dt HdB/dt = V\int _{B}dB H  \eqno{(2)}$$ 
for example, of a macroscopic superconductor cylinder with a radius $R$ and a volume $V = \pi R^{2}L$. The amount of the work (2) performed in the normal state depends on the rate of the magnetic field increase: the work is equal to the energy of the magnetic field $W_{n} = \int _{B=0}^{B=\mu _{0}H_{1}}dB H = \mu _{0}H_{1}^{2}/2 = E_{m}$ at very slow increase and to twice this energy $W_{sn} = \int _{B=0}^{B=\mu _{0}H_{1}}dB H_{1} = \mu _{0}H_{1}^{2} = 2E_{m}$ at a very rapid increase \cite{Hirsch2024}. Half of this work is spent on creating the energy of the magnetic field $E_{m} = \mu _{0}H_{1}^{2}/2$, while the other half, the surplus work $W_{surp} = W_{sn} - E_{m} = \mu _{0}H_{1}^{2}/2$, induces Foucault currents, which dissipate into Joule heat $\Delta ST = W_{surp}$. Therefore the total energy $U = F + ST$ increases on the amount equal to the whole work $\Delta U = W_{sn} = 2E_{m}$, while the free energy $F = U - ST$ increases only on the amount $\Delta F = \Delta U - \Delta ST = W_{sn} - W_{surp} = E_{m}$ equal the energy of the magnetic field which can be used for a negative work. Thus, the increase in free energy of the normal state  
$$F_{nH} = F_{n0} + E_{m} = F_{n0} + \frac{\mu_{0}H^{2}}{2} \eqno{(3)}$$
does not depends on the rate of the magnetic field increase. The power source should perform only a small work  $W_{s} = E_{k}$ equal to the kinetic energy  $E_{k} = \mu _{0}H_{1}^{2}\lambda_{L}/R$ of the surface shielding current $j = (H/\lambda_{L})\exp (r - R)/\lambda_{L}$ \cite{PhysicaC2021} to create a magnetic field $H_{1}$ in the superconducting state when the magnetic flux density $B = \mu _{0}H\exp (r - R)/\lambda_{L}$ penetrates only into a thin surface layer of thickness equal the London penetration depth $\lambda _{L} = (m/\mu _{0}q^{2}n_{s})^{0.5} \approx 50 \ nm = 5 \ 10^{-8} m$. This work is negligibly small $W_{s}/W_{n} = 2\lambda_{L}/R \approx 0.0001$ in the real case of a macroscopic cylinder with a radius $R \approx 1 \ mm = 10^{-3} m$. Therefore we may conclude that magnetic field should not change free energy 
$$F_{sH} = F_{s0} \eqno{(4)}$$
of the macroscopic cylinder in the superconducting state.  

Because of the belief in thermodynamics, no one has noticed for many years that the equality $F_{sH} = F_{nH}$ at $H = H_{c}(T)$ cannot be obtained from the inequality $F_{s0} < F_{n0}$ at $T < T_{c}$ and $H = 0$ if the free energy increases with magnetic field in the normal state and does not change in the superconducting state in accordance with the equations (3) and (4). No one has noticed also that C.J. Gorter and H. Casimir in 1934 used a false claim that "{\it the work, done by the current in the coil, which brings about $H$}" \cite{Gorter1934} creates the energy $dW = -HdM$ of magnetization $M = B - \mu_{0}H$ rather than the energy of magnetic field $dW = HdB$ in order to deduce the equality $F_{sH} = F_{nH}$ at $H = H_{c}(T)$ from the inequality $F_{s0} < F_{n0}$ at $H = 0$. The falsity of this claim is so obvious that it is surprising that this claim could have been done by physicists. According to this claim and contrary to the law of conservation of energy, no work is needed in order to create a magnetic field $H$ in the volume $V$ of an empty coil or solenoid, inside which the magnetization $M = B - \mu_{0}H = 0$. 

Despite this obvious falsity, the authors of most books \cite{Shoenberg1938,Shoenberg1952,Kittel956,Lynton1962,Buckel1972,Schmid1997} following the claim of C.J. Gorter and H. Casimir \cite{Gorter1934} stated that magnetic field should not change free energy in the normal state  
$$F_{nH} = F_{n0} \eqno{(5)}$$
and should increase 
$$F_{sH} = F_{s0} + E_{m} = F_{s0} + \frac{\mu_{0}H^{2}}{2} \eqno{(6)}$$
in the superconducting state. The authors of only a few books, mostly future Nobel prize winners, did not follow the false claim of C.J. Gorter and H. Casimir \cite{Gorter1934}. V.L. Ginzburg \cite{Ginzburg1946} and P.G. de Gennes \cite{Gennes1966} understood that magnetic field should increase free energy of a bulk superconductor in the normal state, in accordance with (3), and should not change in the superconducting state, in accordance with (4).    

\section{Hirsch repeated a mistake made by Gorter and Casimir.}
Jorge Hirsch, unlike the authors of most books \cite{Shoenberg1938,Shoenberg1952,Kittel956,Lynton1962,Buckel1972,Schmid1997} and like V.L. Ginzburg \cite{Ginzburg1946} and P.G. de Gennes \cite{Gennes1966} understands that the power source of the solenoid creates the energy of magnetic field $HB/2$ rather than the energy $-HM/2$ of magnetization $M = B - \mu_{0}H$. He understands also that the same amount of the positive work $W_{ns} = \mu _{0}H_{1}^{2} = 2E_{m}$ is done both during the transition of a bulk superconductor to the normal state at $H = H_{c}(T)$ and during a rapid increases in the magnetic field in the normal state \cite{Hirsch2024}. But J. Hirsch repeated an other mistake made by C.J. Gorter and H. Casimir \cite{Gorter1934}. The authors of most books \cite{Shoenberg1938,Shoenberg1952,Kittel956,Lynton1962,Buckel1972,Schmid1997}, following "{\it the thermodynamics of superconductors developed by Gorter and Casimir}" \cite{Lynton1962}, did not take into account that the equality of free energies     
$$F_{nH} = F_{sH}  \eqno{(7)}$$ 
at $H = H_{c}(T)$ cannot be correct also because of the work $W_{sn} = \mu _{0}H_{1}^{2} = 2E_{m}$ performed at this superconducting transition. 

The work should be done according to both the correct opinion of V.L. Ginzburg \cite{Ginzburg1946} and P.G. de Gennes \cite{Gennes1966} and the false claim of C.J. Gorter and H. Casimir \cite{Gorter1934}, since both the magnetic flux density $B$ and magnetization $M = B - \mu_{0}H$ should change during the superconducting transition at $H = H_{c}(T)$. Only the sign of the work should be different. The work during the transition to the normal state must be positive $W_{sn} = \mu _{0}H_{c}^{2} = 2E_{m}$ according to V.L. Ginzburg \cite{Ginzburg1946} and P.G. de Gennes \cite{Gennes1966} and negative $W_{sn,GC} = -\mu _{0}H_{c}^{2} = -2E_{m}$ according to C.J. Gorter and H. Casimir \cite{Gorter1934}, since the magnetic flux density changes from $B = 0$ to $B = \mu_{0}H_{c}$ and the magnetization changes from $M = -\mu_{0}H_{c}$ to $M = 0$ during this transition. At the transition to the superconducting state, reverse processes occur due to the Meissner effect and the work of opposite sign is done: negative $W_{ns} = -\mu _{0}H_{c}^{2} = -2E_{m}$ according to the correct opinion \cite{Ginzburg1946,Gennes1966} and positive $W_{ns,GC} = \mu _{0}H_{c}^{2} = 2E_{m}$ according to the false claim \cite{Gorter1934}. 

V.L. Ginzburg \cite{Ginzburg1946} and P.G. de Gennes \cite{Gennes1966}, unlike authors of most books \cite{Shoenberg1938,Shoenberg1952,Kittel956,Lynton1962,Buckel1972,Schmid1997}, understood that the equality of free energies (7) at $H = H_{c}(T)$ cannot be deduced without contradiction with the second law of thermodynamics. The negative work $W_{ns} = \int _{B=\mu _{0}H_{c}}^{B=0}dB H_{c} = -\mu _{0}H_{c}^{2}$ increasing the energy of the power source \cite{Hirsch2024} must decrease the total energy $U = F + ST$ of the superconductor $\Delta U = W_{ns} = -\mu _{0}H_{c}^{2}$ according to the law of conservation of energy. This work must decrease the heat $\Delta U = \Delta F + \Delta ST = \Delta ST = W_{ns} = -\mu _{0}H_{c}^{2}$ if the free energy does not change $\Delta F = F_{nH} - F_{sH} = 0$ in accordance with (7), as the authors of most books \cite{Shoenberg1938,Shoenberg1952,Kittel956,Lynton1962,Buckel1972,Schmid1997} were sure. The increase of the energy of the power source by reducing heat $\Delta ST_{1} = -\mu _{0}H_{c}^{2}$ at a constant temperature $T_{1}$ contradicts the second law of thermodynamics, as it implies a decrease in entropy by a macroscopic amount $\Delta S = -\mu _{0}H_{c}^{2}/T_{1}$. 

Because of the belief in the second law of thermodynamics, V.L. Ginzburg \cite{Ginzburg1946} and P.G. de Gennes \cite{Gennes1966} were sure that the work can change only free energy. According to the equation 
$$F_{sH} - F_{nH} = W_{ns} = - 2E_{m} = - \mu _{0}H_{c}^{2}  \eqno{(8)}$$
deduced by V.L. Ginzburg, free energy per unit volume of the superconductor is decreased by the negative work $W_{ns} = -2E_{m} = - \mu _{0}H_{c}^{2}$ perform by Faraday electromagnetic force at the transition from the normal to superconducting state at $H = H_{c}(T)$. P.G. de Gennes deduced the positive work $W_{sn} = 2E_{m} = \mu _{0}H_{c}^{2}$ performed during the reverse transition from the superconducting to normal state \cite{Gennes1966} which increases free energy $F_{nH} - F_{sH} = 2E_{m} = \mu _{0}H_{c}^{2}$. It is surprising that no one has noticed for many years that equation (8) deduced in the books \cite{Ginzburg1946,Gennes1966} is opposite to equation (7) written in most books \cite{Shoenberg1938,Shoenberg1952,Kittel956,Lynton1962,Buckel1972,Schmid1997}. Equally surprising is the fact that C.J. Gorter and H. Casimir \cite{Gorter1934} did not notice that they obtained the equality of free energies (7) at the cost of contradicting the second law of thermodynamics. 

C.J. Gorter and H. Casimir considered  in 1934 a cycle, see figure in \cite{Gorter1934}, which W.H. Keesom considered in the same year: "{\it Suppose we are realizing a cycle as considered by Gorter, 1) Cooling from just above the normal transition point $T_{c}$ to a temperature $T_{2} < T_{0}$, with external field $H = 0$, 2) applying a magnetic field $H_{2}$, just below the threshold value at $T_{2} < T_{c}$, 3) increasing the temperature to $T_{c}$, the field $H_{2}$ being kept constant, and 4) switching off the magnetic field $H = 0$. In passing (in the beginning of process 3) the threshold value curve $H_{c}(T)$, the electromotive force on the solenoid that maintains the magnetic field must do an amount of work equal to twice the energy of the field that comes into existence in the metal}" \cite{Keesom1934}. This cycle is called 'Gorter cycle' by Hirsch in the article \cite{Hirsch2017PRB}. It should be emphasized that the work performed by the power source of the solenoid during the transition of the bulk superconductor to the normal state is positive $W_{sn} = V\int _{0}^{\mu _{0}H}dB H_{c} = V\mu_{0}H_{c}^{2}$ according to W.H. Keesom \cite{Keesom1934} and negative $W_{snGC} = -V\int _{-\mu _{0}H}^{0}dM H_{c} = -V\mu_{0}H_{c}^{2}$ according to C.J. Gorter and H. Casimir \cite{Gorter1934}. Thus, the contradiction between the most books \cite{Shoenberg1938,Lynton1962,Shoenberg1952,Kittel956,Buckel1972,Schmid1997} and the books \cite{Ginzburg1946,Gennes1966} arose back in 1934.

According to the correct opinion W.H. Keesom \cite{Keesom1934}, V.L. Ginzburg \cite{Ginzburg1946}, P.G. de Gennes \cite{Gennes1966} and J. Hirsch \cite{Hirsch2024} the power source of the solenoid performs the positive work $W_{sn} = V\mu_{0}H_{2}^{2} = 2E_{m}$ at $H_{2} = H_{c}(T)$ and the negative work $W_{G4} = -V\mu_{0}H_{2}^{2}/2 = - E_{m}$ during process 4. The total work in the Gorter cycle is positive and equal the surplus work $W_{G} = W_{sn} + W_{G4} = 2E_{m} - E_{m} =  V\mu_{0}H_{2}^{2}/2 = W_{surp}$ according to the correct opinion. According to the false claim of C.J. Gorter and H. Casimir \cite{Gorter1934} the positive work $W_{G2} = -V\int _{0}^{-\mu _{0}H_{2}}dM H = V\mu_{0}H_{2}^{2}/2$ is performed during process 2, when the magnetic field increases from $H = 0$ to $H = H_{2}$ in the superconducting state at $T_{2} < T_{c}$ and the negative work $W_{snGC} = -V\int _{-\mu _{0}H}^{0}dM H_{c} = -V\mu_{0}H_{c}^{2}$ is performed during the transition at $H_{2} = H_{c}(T_{2})$. The total work in the Gorter cycle is also equal the surplus work, but it is negative $W_{G,GC} = W_{G2} + W_{snGC} = E_{m} - 2E_{m} = -V\mu_{0}H_{2}^{2}/2 = -W_{surp}$.       

According to the first law of thermodynamics or the law of conservation of energy the total work in the Gorter cycle should be equal the heat absorbed by the superconductor from the environment. Therefore C.J. Gorter and H. Casimir wrote the following equation   
$$-\int _{T_{2}}^{T_{c}}dT(c_{s} - c_{n}) + Q_{2} = -\int _{0}^{\sigma _{2}}d\sigma H + H_{2}\sigma_{2} \eqno{(9)}$$
(equation (11) in their article \cite{Gorter1934}), where the index 2 indicates the transition point at $H_{2} = H_{c}(T_{2})$ and $Q_{2}$ is the heat of transition at $H_{2} = H_{c}(T_{2})$. $c_{s}$ and $c_{n}$ are the heat capacities in the superconducting and normal states respectively; $\int _{T_{2}}^{T_{c}}dT c_{s}$ is the heat released by the superconductor to the environment in process 1; $\int _{T_{2}}^{T_{c}}dT c_{n}$ is the heat absorbed by the superconductor from the environment in process 3; $\sigma  = VM = - V\mu_{0}H$ is the magnetic moment in superconducting state at a magnetic field $H$; $\sigma _{2} = VM = - V\mu_{0}H_{2}$ is the same at $H_{2} = H_{c}(T_{2})$; $- \int _{0}^{\sigma _{2}}d\sigma H = W_{G2} = V\mu_{0}H_{2}^{2}/2$ is the positive work performed in process 2; $H_{2}\sigma_{2} = W_{snGC} = -V\mu_{0}H_{c}^{2}$ is the negative work performed during the transition to the normal state at $H_{2} = H_{c}(T)$. 

C.J. Gorter and H. Casimir did not notice that their equation (9) (equation (11) in \cite{Gorter1934}) contradicts the second law of thermodynamics because they forgot that the first law of thermodynamics must be valid not only for a closed cycle, but for any process, for example, for the superconducting transition at $H_{2} = H_{c}(T_{2})$. Not only equation (9) but also the equation  
$$Q_{2} = H_{2}\sigma_{2} = W_{snGC} = -V\mu_{0}H_{c}^{2} \eqno{(10)}$$
must be valid according to this fundamental law. The equality of free energies (7) can be deduced from the equation (10): $F_{nH} - F_{sH} = \Delta F = \Delta U - \Delta ST_{2} = W_{snGC} - Q_{2} = -V\mu_{0}H_{c}^{2} - (-V\mu_{0}H_{c}^{2}) = 0$. But the equation (10) contradicts obviously the second law of thermodynamics since it implies the decrease in total entropy $\Delta S = -V\mu_{0}H_{c}^{2}/T_{2}$ at constant temperature $T_{2}$. 

J. Hirsch did not notice also that his equation (10) in \cite{Hirsch2024} contradicts the second law of thermodynamics since he, like C.J. Gorter and H. Casimir \cite{Gorter1934}, forgot that the law of conservation of energy must be valid not only for a closed cycle. Equation (10) in \cite{Hirsch2024} should repeat equation (9) (equation (11) in \cite{Gorter1934}) since it is also deduced on the base of the law of conservation of energy in the Gorter cycle: the total work in this cycle $W_{aG} = W_{aG1} + W_{ns} = \int _{B=0}^{B=\mu _{0}H_{2}}dB H - V\mu_{0}H_{2}^{2} =   -V\mu_{0}H_{2}^{2}/2$ should be equal 
$$\int _{T_{2}}^{T_{c}}dT(c_{s} - c_{n}) - Q_{2} = V\int _{B=0}^{B=\mu _{0}H_{2}}dB H - V\mu_{0}H_{2}^{2} \eqno{(11)}$$
the heat absorbed by the superconductor from the environment $Q_{1} - Q_{2} - L(T_{2}) $ according to \cite{Hirsch2024} and the same amount $\int _{T_{2}}^{T_{c}}dT(c_{s} - c_{n}) - Q_{2}$ according to \cite{Gorter1934}.  

J. Hirsch considered the Gorter cycle counterclockwise \cite{Hirsch2024}, not clockwise as C.J. Gorter and H. Casimir did \cite{Gorter1934}. Therefore the heat absorbed from the environment should have the opposite sign, see (9) and (11). $Q_{1} =\int _{T_{2}}^{T_{c}}dT c_{s}$ is the heat absorbed by the system from the environment in process 4 according to equation (2) and $Q_{2} = \int _{T_{2}}^{T_{c}}dT c_{n} $ is the heat released by the system to the environment in process 2 according to equation (3) of the Hirsch article \cite{Hirsch2024}. Hirsch denoted the heat of transition $Q_{2}(T_{2})$ as $L(T_{2})$ \cite{Hirsch2024}. But the work $W_{ns} = - V\mu_{0}H_{2}^{2}$ performed at $H_{2} = H_{c}(T_{2})$ and in the cycle $W_{aG} = -V\mu_{0}H_{2}^{2}/2$ must be negative, as in the article \cite{Gorter1934}, since Hirsch did not follow the false claim of C.J. Gorter and H. Casimir. $W_{aG1} = \int _{B=0}^{B=\mu _{0}H_{2}}dB H = \mu _{0}H_{2}^{2}/2 = E_{m}$ is the positive work performed in process 1 of the counterclockwise Gorter cycle when magnetic field $H$ is increased at $T = T_{c}$ from $H = 0$ to $H = H_{2}$ in the normal state. We use the notations $T_{2}$ and $H_{2}$ of C.J. Gorter and H. Casimir \cite{Gorter1934} instead of the notations $T_{1}$ and $H_{1}$ used by Hirsch \cite{Hirsch2024} for the same parameters. 

J. Hirsch made obvious mistake contradicting to the law of conservation of energy in equation (10) of his article \cite{Hirsch2024} since the negative work $W_{aG} = -V\mu _{0}H_{2}^{2}/2$ performed in the Gorter cycle counterclockwise increases the energy of the power source according to his article \cite{Hirsch2024} and therefore must decrease rather than increase heat energy. The equation (11), like equation (9) deduced by C.J. Gorter and H. Casimir \cite{Gorter1934}, contradicts to the second law of thermodynamics since the law of conservation of energy must be valid not only in the Gorter cycle but also during the transition at $H_{2} = H_{c}(T_{2})$, and therefore the equation $-Q_{2} = -V\mu_{0}H_{c}^{2}$ must be valid. Despite this contradiction, C.J. Gorter and H. Casimir assumed \cite{Gorter1934}, and J. Hirsch even had the illusion he could prove that the second law of thermodynamics is not violated in the Gorter cycle \cite{Hirsch2024}.    

C.J. Gorter and H. Casimir understood that the validity of the second law of thermodynamics can be proved for any process of the Gorter cycle except the transition at $H_{2} = H_{c}(T_{2})$. Therefore they wrote: "{\it Up to now, we have hardly made any special hypothesis, but now we introduce the assumption, that the second law of thermodynamics applies also to the transition}" \cite{Gorter1934}. According to this assumption the entropy remains unchanged in the Gorter cycle $-\int _{T_{2}}^{T_{c}}dT c_{s}/T + Q_{2}/T_{2} + \int _{T_{2}}^{T_{c}}dT c_{n}/T  = 0$. C.J. Gorter and H. Casimir  have deduced the equation $Q_{2} = T_{2}(S_{n}(T_{2}) - S_{s}(T_{2}))$ for the heat of transition $Q_{2}$ from this immutability, which J. Hirsch wrote as equation (1) in the beginning of his article \cite{Hirsch2024} without any assumption: $Q_{2} = T_{2}\int _{T_{2}}^{T_{c}}dT c_{s}/T - T_{2}\int _{T_{2}}^{T_{c}}dT c_{s}/T = T_{2}(S_{s}(T_{c}) - S_{s}(T_{2})) - T_{2}(S_{n}(T_{c}) - S_{n}(T_{2})) = T_{2}(S_{n}(T_{2}) - S_{s}(T_{2}))$ since the entropy of normal and superconducting state equal $S_{n}(T_{c}) = S_{s}(T_{c})$ at $T = T_{c}$. 

This equation, written without any assumption or justification, has provoked illusion of J. Hirsch that he could prove that the total entropy does not change in the Gorter cycle, without the assumption made by C.J. Gorter and H. Casimir \cite{Gorter1934}, see equations (16 - 18) in  \cite{Hirsch2024}. J. Hirsch write after equations (18): "{\it Therefore, neither the environment nor the sample changed their entropy in the cycle, consistent with the fact that it was a reversible process. The second law of thermodynamics is satisfied, not violated, contrary to the claim in Ref. \cite{Entropy2022}}" \cite{Hirsch2024}. Thus C.J. Gorter and H. Casimir postulated the contradiction of the transition at $H_{2} = H_{c}(T_{2})$ with the second law of thermodynamics and assumed that this law applies to this transition \cite{Gorter1934}. J. Hirsch went even further than C.J. Gorter and H. Casimir. Having postulated the violation of the second law of thermodynamics, J. Hirsch 'proved' that this law is not violated using equation (1) which C.J. Gorter and H. Casimir have derived using the assumption that the second law of thermodynamics is valid at $H_{2} = H_{c}(T_{2})$ \cite{Gorter1934}. 

C.J. Gorter and H. Casimir used the equation $Q_{2} = T_{2}(S_{n}(T_{2}) - S_{s}(T_{2}))$ to derive "{\it the basic equation of the thermodynamics of superconductors developed by Gorter and Casimir}" \cite{Lynton1962} from equation (9) (equation (11) in \cite{Gorter1934}). The left part of equation (9) is the free energy difference $-\int _{T_{2}}^{T_{c}}dT(c_{s} - c_{n}) + Q_{2} = U_{s}(T_{2}) - U_{n}(T_{2}) + T_{2}S_{n}(T_{2}) - T_{2}S_{s}(T_{2}) = F_{s0}(T_{2}) - F_{n0}(T_{2})$ in zero magnetic field $H = 0$ at temperature $T_{2}$, since $-\int _{T_{2}}^{T_{c}}dT(c_{s} - c_{n}) = U_{s}(T_{2}) - U_{s}(T_{c}) + U_{n}(T_{c}) - U_{n}(T_{2}) = U_{s}(T_{2}) - U_{n}(T_{2})$ is the difference of total energies at $T_{2}$ because of the equality $U_{s}(T_{c}) = U_{n}(T_{c})$. The right part of equation (9) is the negative surplus work $W_{surpGC} = -\mu _{0}H_{2}^{2}/2$ and therefore 
$$F_{n0} - F_{s0} = \frac{\mu_{0}H_{c}^{2}}{2} \eqno{(12)}$$ 
the difference of free energies between normal and superconducting states $F_{n0} - F_{s0}$ at $T_{2} < T_{c}$ equal the energy of magnetic field in normal state at $H_{2} = H_{c}(T_{2})$. 

C.J. Gorter and H. Casimir derived equation (12) based on the assumption of validity of the second law of thermodynamics from equation (9) that contradicts the second law of thermodynamics and the law of conservation of energy. Despite these contradictions with fundamental laws of physics and logic "{\it the basic equation of the thermodynamics of superconductors developed by Gorter and Casimir}" (12) played an extremely important role in the history of superconductivity research. V.L. Ginzburg and L.D. Landau used the equation (12) in their theory \cite{GL1950}, and J. Bardeen, L.N. Cooper and J.R. Schrieffer successfully explained \cite{BCS1957} why at formation of Cooper pairs the energy decreases exactly by the amount of the magnetic field energy $E_{m} = \mu _{0}H_{c}^{2}/2$ in the normal state at $H_{2} = H_{c}(T_{2})$ in accordance with (12). Everyone believed in '{\it the basic equation of superconductor thermodynamics developed by Gorter and Casimir}' (12) because it is consistent with the common belief that the transition to the superconducting state occurs when its free energy becomes less than the free energy of the normal state. According to equations (5) and (6), written in most books \cite{Shoenberg1938,Shoenberg1952,Kittel956,Lynton1962,Buckel1972,Schmid1997}, the difference of free energies (12) decreases with magnetic field $F_{nH} - F_{sH} = \mu _{0}H_{c}^{2}/2 - \mu _{0}H^{2}/2$ and the transition to the normal state occurs $H_{2} = H_{c}(T_{2})$ when the free energies are equal $F_{nH}(H_{c}) = F_{sH}(H_{c})$ in accordance with (7).  

M. Tinkham \cite{Tinkham1996}, unlike the authors of most books \cite{Shoenberg1938,Shoenberg1952,Kittel956,Lynton1962,Buckel1972,Schmid1997}, did not follow the false claim of C.J. Gorter and H. Casimir \cite{Gorter1934}. But he believed in their equation (12) and derived the free energy change at the superconducting transition at $H_{2} = H_{c}(T_{2})$ (8) from this equation and correct equations (3,4): $F_{nH} - F_{sH} = F_{n0} + \mu _{0}H_{c}^{2}/2 - F_{s0} = \mu _{0}H_{c}^{2}/2 + \mu _{0}H_{c}^{2}/2 = \mu _{0}H_{c}^{2}$. "{\it The basic equation of the thermodynamics of superconductors developed by Gorter and Casimir}" (12) can be derived both from the false equations (5 - 7), written in most books \cite{Shoenberg1938,Shoenberg1952,Kittel956,Lynton1962,Buckel1972,Schmid1997}, and from equations (3,4,8) derived by V.L. Ginzburg \cite{Ginzburg1946} and P.G. de Gennes \cite{Gennes1966} in order to avoid the contradiction with the law of conservation of energy and the second law of thermodynamics. This agreement from the disagreement became possible because of the accidental fact that the work $W_{sn} =  \mu_{0}H_{c}^{2}$ is twice the energy of the magnetic field in the normal state $E_{m} = \mu_{0}H_{c}^{2}/2$: $F_{n0} - F_{s0} = F_{nH} - (F_{sH} - E_{m}) = E_{m} = \mu_{0}H_{c}^{2}/2$ according to (5-7); $F_{n0} - F_{s0} = (F_{nH} - E_{m}) - F_{sH} = 2E_{m} - E_{m} = E_{m} = \mu_{0}H_{c}^{2}/2$ according to (3,4,8). 

J. Hitsch, in contrast to V.L. Ginzburg \cite{Ginzburg1946} and P.G. de Gennes \cite{Gennes1966}, was not able to derive  "{\it the basic equation of the thermodynamics of superconductors developed by Gorter and Casimir}", since he, like C.J. Gorter and H. Casimir \cite{Gorter1934}, did not avoid the contradiction with the second law of thermodynamics. The free energies must be equal $F_{nH}(H_{c}) = F_{sH}(H_{c})$ according to his equation (10) and the law of conservation of energy. J. Hirsch must obtain from $F_{nH}(H_{c}) = F_{sH}(H_{c})$ and correct equations (3,4) the equation $F_{n0} - F_{s0} = -\mu_{0}H_{c}^{2}/2$ rather than equation (12). This equation is indeed derived from equation (11) the left part of which is $F_{n0}(T_{2}) - F_{s0}(T_{2})$, unlike equation (9) $F_{s0}(T_{2}) - F_{n0}(T_{2})$. The opposite of the signs is a consequence of the fact that C.J. Gorter and H. Casimir \cite{Gorter1934} considered the Gorter cycle clockwise, and J. Hirsch \cite{Hirsch2024} counterclockwise. But the work in right part should be negative both in equation (9) and (11) since J. Hirsch \cite{Hirsch2024} did not follow the false claim of C.J. Gorter and H. Casimir \cite{Gorter1934}. J. Hitsch has derived equation (13) in his article \cite{Hirsch2024}, which is "{\it the basic equation of the thermodynamics of superconductors developed by Gorter and Casimir}" (12), only due to his false claim that the negative surplus work $W_{surp-} = -\mu _{0}H_{c}^{2}/2$ is positive in his equation (10). Hirsch had to derive the equation $F_{s0} - F_{n0} = \mu_{0}H_{c}^{2}/2$ rather than (12) because he repeated the mistake of Gorter and Casimir but did not follow their false claim. 

\section{Arbitrariness in definitions is not acceptable in science.}
Many of the mistakes made in the article \cite{Hirsch2024} would have been impossible if J. Hirsch had paid attention to the contradictions between books on superconductivity and read carefully the article of C.J. Gorter and H. Casimir \cite{Gorter1934}. The contradictions between books are the result of neglecting one of the main principles of science: arbitrariness in definitions is not acceptable in science. V.L. Ginzburg \cite{Ginzburg1946}, who was the first to derive in 1946 equations (3,4,8), did not pay attention to the fact that D. Shoenberg used the opposite equations (5-7) in the book \cite{Shoenberg1938}, published in 1938 and even translated into Russian. V.L. Ginzburg understood that "{\it As is known from thermodynamics, the difference in free energies after the transition and before it $F_{sH} - F_{nH}$ is equal to the work done on the body}" \cite{Ginzburg1946}. Contrary to V.L. Ginzburg, D. Shoenberg claimed that the free energy should not change at this transition \cite{Shoenberg1938}. D. Shoenberg was apparently the first who wrote about the Gibbs free energy \cite{Shoenberg1938}. The authors of many books \cite{Shoenberg1938,Shoenberg1952,Kittel956,Lynton1962,Buckel1972} were sure that equations (5-7) describe the Gibbs free energy although neither Gibss free energy nor any other free energy can increase when no work is done on a body, contrary to (6), and must increase when work is done, contrary to (5).        

M. Tinkham \cite{Tinkham1996}, in contrast to the authors  \cite{Shoenberg1938,Shoenberg1952,Kittel956,Lynton1962,Buckel1972}, did not follow to the false claim of C.J. Gorter and H. Casimir \cite{Gorter1934}. But he used also the name 'Gibbs free energy' in order to create an illusion a possibility of the equality of free energy (7), that must be at the phase transition. He called the free energy described by the equation (3,4,8) as 'Helmholtz free energy' and was sure that "{\it The reason for the awkward necessity of considering the energy of the source of the field is that we carried out the preceding discussion in terms of the Helmholtz free energy}" \cite{Tinkham1996}. He was sure that "{\it The appropriate thermodynamic potential for the case of constant $H$ is the Gibbs free energy $G$. This differs from $F$ by the term $-BH$, which essentially accounts automatically for the work done by the generator. Thus, we consider the Gibbs free energy density}" \cite{Tinkham1996}. The Gibbs free energy density equals  
$$G = F - BH \eqno{(13)}$$
according to M. Tinkham \cite{Tinkham1996}.

It must be emphasized that 'Gibbs free energy' of M. Tinkham \cite{Tinkham1996} differs from 'Gibbs free energy' of D. Shoenberg, used in the most books \cite{Shoenberg1938,Shoenberg1952,Kittel956,Lynton1962,Buckel1972}: the 'Gibbs free energy' of D. Shoenberg is not changed in magnetic field in the normal state (5) and increases in the superconducting state (6), while the 'Gibbs free energy' of M. Tinkham \cite{Tinkham1996} does not change in the superconducting state, in accordance with the equation (4), and decreases in the normal state 
$$G_{nH} = G_{n0} - E_{m} = G_{n0} - \frac{\mu_{0}H^{2}}{2} \eqno{(14)}$$
on a amount equal the magnetic field energy in the normal state $E_{m} = \mu _{0}H^{2}/2$. A.A. Abrikosov \cite{Abrikosov1988} used the same trick as M. Tinkham \cite{Tinkham1996} to get the equality (7), from the equation (8). But A.A. Abrikosov did not try to justify this trick by referring to the difference between Helmholtz and Gibbs free energies, apparently realizing that it does not make sense in this case. 

It is surprising that the future Nobel prize winner did not notice that his trick provoked the obvious contradiction with the law of energy conservation in equation (14), derived also in his book \cite{Abrikosov1988}. The energy of the magnetic field $E_{m} = \mu_{0}H^{2}/2$ can reduce no free energy, since any free energy $F = U - ST$ is the energy capable to perform  a work without thermal energy $ST$. The energy of the magnetic field $E_{m} = \mu_{0}H^{2}/2$ is capable to perform a work without thermal energy and therefore it is a part of any free energy. The incorrect equation (14) derived by M. Tinkham \cite{Tinkham1996} and A.A. Abrikosov \cite{Abrikosov1988} differs from the correct equation (3) derived by V.L. Ginzburg \cite{Ginzburg1946} and P.G. de Gennes \cite{Gennes1966} by a sign before the energy of the magnetic field $E_{m} = \mu_{0}H^{2}/2$ because an amount of the positive $W_{sn} = \mu _{0}H_{c}^{2}$ \cite{Gennes1966} and negative $W_{ns} = -\mu _{0}H_{c}^{2}$ \cite{Ginzburg1946} work performed at $H = H_{c}$ "{\it equal to twice the energy of the field that comes into existence in the metal}" \cite{Keesom1934}.

The three options of the free energy in the normal state demonstrate once again that arbitrariness in definitions must not be acceptable in science. In this case, the issue is most obvious, since the work done to create a magnetic field in the volume of the superconductor in the normal state should not differ from the work done in an empty solenoid if the magnetic field increases slowly enough. Despite the obviousness of this issue the authors of the most books \cite{Shoenberg1938,Shoenberg1952,Kittel956,Lynton1962,Buckel1972,Schmid1997} were sure that free energy is not changed in magnetic field in accordance with equation (5) written in their books. In order to get equality of free energies (7) M. Tinkham and A.A. Abrikosov made an even more amazing assertion that free energy can decrease in accordance with equation (14), deduced in their books \cite{Tinkham1996,Abrikosov1988}.     

It is surprising that only V.L. Ginzburg \cite{Ginzburg1946} and P.G. de Gennes \cite{Gennes1966} did not forget that the power source of the solenoid creating a magnetic field $H$ increases the free energy on $E_{m} = \mu_{0}H^{2}/2$, in accordance with equation (3). But they must have forgotten that free energy cannot change at a phase transition, as at any reversible thermodynamic process. V.L. Ginzburg \cite{Ginzburg1946} and P.G. de Gennes \cite{Gennes1966} did not simply share the common belief that the superconducting transition in the critical magnetic field $H_{2} = H_{c}(T_{2})$ is a first order phase transition. They derived the equations for the jump in entropy $S_{n} - S_{s} = -\mu_{0}H_{c}dH_{c}/dT$ and for the latent heat 
$$L(T_{2}) = T_{2}(S_{n} - S_{s}) = -T_{2}\mu_{0}H_{c}\frac{dH_{c}}{dT}  \eqno{(15)}$$     
derived in the most books \cite{Shoenberg1938,Shoenberg1952,Kittel956,Lynton1962,Buckel1972,Schmid1997}. But these equations can be derived mathematically only from equation (7) but not from equation (8), since entropy is equal to the first derivative of free energy with respect to temperature $S = - dF/dT$ and, consequently, the jump in entropy can be finite only if the temperature dependence is not changed by jump at $T = T_{2}$ and $H_{2} = H_{c}(T_{2})$. 

The equation (15) can be derived from the false equations (5-6) written in the most books \cite{Shoenberg1938,Shoenberg1952,Kittel956,Lynton1962,Buckel1972,Schmid1997} $dF_{nH}/dT = dF_{n0}/dT = -S_{n} = dF_{sH}/dT = dF_{s0}/dT + \mu_{0}H_{c}dH_{c}/dT = -S_{s} + \mu_{0}H_{c}dH_{c}/dT $ and also from the false equations (3,7,14) obtained by M. Tinkham \cite{Tinkham1996} and A.A. Abrikosov \cite{Abrikosov1988} $dG_{nH}/dT = dG_{n0}/dT - \mu_{0}H_{c}dH_{c}/dT = -S_{n} - \mu_{0}H_{c}dH_{c}/dT = dG_{sH}/dT = dG_{s0}/dT = -S_{s}$ since free energies of two phases are equal according to equation (7) at $T = T_{2}$ and $H_{2} = H_{c}(T_{2})$ and therefore the first derivative of free energy with respect to temperature $S = - dF/dT$ can be finite. But the first derivative must be infinite according to equation (8): $(F(T_{2} + dT/2) - F(T_{2} - dT/2))/dT = (F_{nH} - F_{sH})/dT = \mu_{0}H_{c}^{2}/dT = \infty $ at $dT \rightarrow 0$. V.L. Ginzburg \cite{Ginzburg1946} and P.G. de Gennes \cite{Gennes1966} derived the equations for the jump in entropy and the latent heat (15) from "{\it the basic equation of the thermodynamics of superconductors developed by Gorter and Casimir}" (12). This deduction is mathematically incorrect because the difference of free energies on the left side of equation (12) is a function of $T = T_{2}$ and $H = 0$, while the magnetic field energy on the right side of (12) is a function of $T = T_{2}$ and $H_{2} = H_{c}(T_{2})$. 

Following the common misconception, J. Hirsch claims at the beginning of the article \cite{Hirsch2024} that the heat of transition at $H_{2} = H_{c}(T_{2})$ ($Q_{2}$ in \cite{Gorter1934}) is the latent heat $L(T_{2}) = T_{2}(S_{n}(T_{2}) - S_{s}(T_{2}))$. That is, he claims at the beginning of the article what he intends to prove in this article: the second law of thermodynamics is satisfied, not violated \cite{Hirsch2024}. The latent heat can be observed only at a first order phase transition, which satisfies the second law of thermodynamics by definition. But the latent heat, which compensates $\Delta F = \Delta U - L = \Delta U - T(S_{n} - S_{s}) = 0$ the change in total energy $\Delta U $ and leaves the free energy unchanged $F_{sH} = F_{nH}$, cannot perform work, also by definition. Therefore C.J. Gorter and H. Casimir did not claim that the heat of transition $Q_{2}$ is the latent heat and used the assumption, that the second law of thermodynamics applies also to the transition \cite{Gorter1934}. The claim that the heat of transition can be the latent heat is false also because the latent heat equal $L(T_{2}) = \mu_{0}H_{c}^{2}$ according to equation (10) in the Hirsch article \cite{Hirsch2024} differs from the latent heat (15), written in all books.     

J. Hirsch \cite{Hirsch2024} followed to C.J. Gorter and H. Casimir \cite{Gorter1934} in the main. But he did not realize that his equation  (13) in \cite{Hirsch2024} is "{\it the basic equation of the thermodynamics of superconductors developed by Gorter and Casimir}" (12) rather than the change in free energy during the superconducting transition at $H_{2} = H_{c}(T_{2})$. Therefore he used the trick of M. Tinkham \cite{Tinkham1996}: "{\it The reader may wonder how Eq. (13) is consistent with the expectation that in a first order phase transition at coexistence the free energies of the two phases should be equal. That is simply a matter of definition of free energies, as explained e.g. by Tinkham \cite{Tinkham1996}. The Helmholtz free energy defined by Eq. (11) is not the same for the two phases because it does not include the energy of the current source that keeps the external magnetic field constant, while an appropriately defined Gibbs free energy is \cite{Tinkham1996}}" \cite{Hirsch2024}. This justification shows that J. Hirsch doesn't quite understand what he is deducing: according to his equation (10) in \cite{Hirsch2024}, like equation (11) in \cite{Gorter1934}, the negative work $W_{ns} = -\mu_{0}H_{c}^{2}$ reduces only heat and does not change the free energy $F_{nH} = F_{sH}$ in accordance with equality (7).  
        
\section{Regress in the understanding of thermodynamics.}
The numerous mistakes made by J. Hirsch in the article \cite{Hirsch2024} indicate a regression in the understanding of thermodynamics that was provoked by the belief in thermodynamics. The regression began a century ago when W.H. Keesom (1924 Rapp. 4-A Congr. Phys. Solvay, 288) derived  equation (15) for the latent heat, see the article \cite{Keesom1934f}. This unreasonable belief that the superconducting transition at $H = H_{c}(T)$ is a first order phase transition has prevailed for a hundred years despite the work $W_{sn} = \mu _{0}H_{c}^{2}$ performed in accordance with the equation (2). Because of this belief, everyone ignores the fact that the work changing $\Delta U = \mu _{0}H_{c}^{2}$ the total energy $U = F + ST$, according to the law of conservation of energy, must change either free energy $\Delta F = \mu _{0}H_{c}^{2}$ or heat $\Delta S T = \mu _{0}H_{c}^{2}$. The unreasonable belief provoked three logically possible misconceptions in books on superconductivity: 1) V.L. Ginzburg \cite{Ginzburg1946} and P.G. de Gennes \cite{Gennes1966} had to forget that free energy cannot change during a phase transition; 2) M. Tinkham \cite{Tinkham1996} and A.A. Abrikosov \cite{Abrikosov1988} had to contradicts to the law of conservation of energy by the claim that the work $W_{sn} = \mu _{0}H_{c}^{2}$ should not change either free energy or heat $\Delta U = \Delta F + \Delta ST = 0$; 3) C.J. Gorter and H. Casimir \cite{Gorter1934}, and the authors of most books \cite{Shoenberg1938,Shoenberg1952,Kittel956,Lynton1962,Buckel1972,Schmid1997} had to contradict the second law of thermodynamics by the claim that the work changes heat $\mu _{0}H_{c}^{2} = \Delta S T$. 

Various misconceptions appeared as early as 1934. W.H. Keesom, in contrast to C.J. Gorter and H. Casimir \cite{Gorter1934}, understood that the power source of the solenoid creates magnetic field energy $BH/2$ rather than the energy $-MH/2$ of magnetization $M = B - \mu_{0}H$. His belief that the transition of a bulk superconductor to the normal state at $H = H_{c}(T)$ is a first order phase transition is particularly surprising, since this belief was false for at least four reasons according to his knowledge before 1933: 

1) W.H. Keesom knew about the work $W_{sn} = \mu _{0}H_{c}^{2}$ performed at $H = H_{c}(T)$; 

2) W.H. Keesom understood that the half of the work $W_{sn} = \mu_{0}H_{c}^{2}$ creates the energy of the magnetic field $E_{m} = \mu_{0}H_{c}^{2}/2$, which is part of free energy. Because of the change in the free energy
$$F_{nH} = F_{sH} + E_{m} = F_{sH} + \frac{\mu_{0}H_{c}^{2}}{2} \eqno{(16)}$$
the transition cannot be a phase transition; 

3) W.H. Keesom imagined before the discovery of the Meissner effect that the second half of the work, the surplus work  $W_{surp} = W_{sn} - E_{m} = \mu _{0}H_{c}^{2}/2$, generates Joule heat: "{\it Till now we imagined that the surplus work served to deliver the Joule-heat developed by the persistent currents the metal getting resistance while passing to the non-supraconductive condition}" \cite{Keesom1934}; 

4) W.H. Keesom should have known until 1933 that the transition cannot be reversible since the magnetic flux cannot be pushed out of a perfect conductor in accordance with physical laws known in 1933, such as Faraday's law and the law of angular momentum conservation. 

The Meissner effect could eliminate only the fourth reason, and only experimentally, not in theory. It is surprising that E.A. Lynton wrote in the book \cite{Lynton1962} published in 1962: "{\it Gorter actually predicted the Meissner effect, pointing out that the success of the thermodynamic approach strongly supports the reversibility of the superconducting transition}". How could Gorter or anyone else predict the Meissner effect, which contradicts the laws of physics known in 1933? The attitude of E.A. Lynton \cite{Lynton1962} to this contradiction expresses the attitude of almost all experts on superconductivity. This contradiction was ignored for many years, until J. Hirsch began to pay attention to it. J. Hirsch expressed surprise in 2010 about the ignoring of the obvious contradiction: 

"{\it Strangely, the question of what is the force propelling the mobile charge carriers and the ions in the superconductor to move in direction opposite to the electromagnetic force in the Meissner effect was essentially never raised nor answered to my knowledge, except for the following instances: \cite{LondonH1935} (H. London states: 'The generation of current in the part which becomes supraconductive takes place without any assistance of an electric field and is only due to forces which come from the decrease of the free energy caused by the phase transformation,' but does not discuss the nature of these forces), \cite{PRB2001} (A.V. Nikulov introduces a 'quantum force' to explain the Little-Parks effect in superconductors and proposes that it also explains the Meissner effect)}" \cite{Hirsch10Meis}. 
 
J. Hirsch compares in 2020 the attitude of most experts to this contradiction with the attitude of the characters of the fairy tale "The Emperor's New Clothes" by Hans Christian Andersen to the emperor's new clothes: 

"{\it Heaven help me he thought as his eyes flew wide open, 'I can't see anything at all'. But he did not say so.}"

"{\it 'Heaven help me', thought smart students that couldn't understand how BCS theory explains the Meissner effect. 'I can't possibly see how momentum conservation is accounted for and Faraday's law is not violated'. But they did not say so}" \cite{HirschAPS}. 

W.H. Keesom could not understand in 1933 why the magnetic flux can be pushed out of a bulk superconductor. But he understood that the Meissner effect violates the second law of thermodynamics if the surplus work generates Joule heat, as he imagined until 1933. Belief in this law caused W.H. Keesom to change his opinion about the surplus work: "{\it As, however, the conception of Joule-heat can rather difficult be reconciled with reversibility we think now that there must be going on another process that absorbs energy}" \cite{Keesom1934}. It should be emphasized that only faith in the second law of thermodynamics, but not experimental results, forced W.H. Keesom to change his opinion. J. Hirsch is sure \cite{Hirsch2024} in "{\it the fact that NO Joule heat is generated in the process where the superconductor in a magnetic field makes the transition to the normal state has been experimentally verified through numerous experiments by W.H. Keesom and coworkers \cite{Keesom1934,Keesom1934f}}". But results of experiments made by W.H. Keesom and coworkers before 1933 when W.H. Keesom "{\it imagined that the surplus work served to deliver the Joule-heat}" \cite{Keesom1934} do not differ from results of experiments made after 1933 when W.H. Keesom began to think "{\it that there must be going on another process that absorbs energy}" \cite{Keesom1934}. 

It is written in the articles \cite{Keesom1934,Keesom1934f} that results of measurements which Hirsch considers as experimental evidence that no Joule heat is generated were obtained in 1932 when W.H. Keesom did not question Joule heat. The experimental results did not change after 1933, but their interpretation changed. W.H. Keesom was sure after 1933 that he measures the latent heat \cite{Keesom1934f} since the concept of latent heat, unlike the concept of Joule heat, is easy to reconciled with reversibility. The latent heat can be observed only during the first-order phase transition, which is a reversible process by definition. C.J. Gorter and H. Casimir, unlike J. Hirsch \cite{Hirsch2024}, did not claim that the heat of transition is the latent heat and assumed rather than proved "{\it that the second law of thermodynamics applies also to the transition}" \cite{Gorter1934} since they understood that the latent heat cannot be distinguished from other types of heat experimentally. 
   
The latent heat cannot be observed at $H = H_{c}(T)$ since this transition cannot be a first order phase transition because of at least three reasons listed above. W.H. Keesom set the task to eliminate the third reason, which had not been solved, but he ignored the first and second reasons. He ignored that the free energy must change during the transition at $H_{2} = H_{c}(T_{2})$ in accordance with equation (16). Ignoring the first reason provoked three various misconceptions. According to equality (7), written in most books \cite{Shoenberg1938,Shoenberg1952,Kittel956,Lynton1962,Buckel1972,Schmid1997} and also in books \cite{Tinkham1996,Abrikosov1988}, the energy of magnetic field should not change when the magnetic flux $\Phi = \pi R^{2}\mu_{0}H_{c}$ penetrates in the volume of the cylinder. Unlike this misconception, not only the energy of magnetic field $E_{m} = \mu_{0}H_{c}^{2}/2$ but also the surplus work  $W_{surp} = W_{sn} - E_{m} = \mu _{0}H_{c}^{2}/2$ increase the free energy according to equation (8) deduced in the books \cite{Ginzburg1946,Gennes1966} because of the faith in the second law of thermodynamics.      

W.H. Keesom did not understand for what purpose the surplus work serves: "{\it It is not clear for what purpose the surplus of absorbed energy serves}" \cite{Keesom1934}. According to  V.L. Ginzburg \cite{Ginzburg1946} and P.G. de Gennes \cite{Gennes1966} the surplus work increases free energy $F_{n0} - F_{s0} = W_{surp} = \mu _{0}H_{c}^{2}/2$ in accordance with "{\it the basic equation of the thermodynamics of superconductors developed by Gorter and Casimir}" (12). J. Hirsch uses the basic equation (12) in order to prove that "{\it NO energy is dissipated as Joule heat}": "{\it Now from Eq. (13) we know that the system requires all that energy to transition into the normal state}" \cite{Hirsch2024}. Of course, the surplus work cannot generate Joule heat or any other heat if this work is required to increase the free energy. But it is surprising that neither Hirsch nor anyone before him, starting with W.H. Keesom, paid attention to the fact that the surplus work is performed in the Gorter cycle, while free energy cannot change in a closed cycle by definition. 

Only the first half of the positive work $W_{sn} = \mu _{0}H_{2}^{2}$, which creates the magnetic field energy $E_{m} = \mu_{0}H_{c}^{2}/2$, returns to the power source with the negative work $W_{G4} = -\mu _{0}H_{2}^{2}/2 = -E_{m}$ during process 4 of the clockwise Gorter cycle. While the positive surplus work $W_{surp+} = W_{sn} + W_{G4} =  2E_{m} - E_{m} = \mu _{0}H_{2}^{2}/2$ should increase $\Delta U = W_{surp+} = V\mu _{0}H_{c}^{2}/2$ the total energy $U = F + ST$ and decrease the energy of the power source in each clockwise Gorter cycle. This work must increase heat $\Delta ST = W_{surp+} = V\mu _{0}H_{c}^{2}/2$ since the free energy cannot change in a closed cycle $\Delta F = 0$. Thus, only the opinion of W.H. Keesom and other physicists, which was before the discovery of the Meissner effect, can be consistent with the basics of thermodynamics: half of the work performed by the power source creates magnetic field energy, while the surplus work generates Joule heat. W.H. Keesom and other physicists could have made this obvious conclusion back in 1933 if their faith in thermodynamics had not been so blind. 

Their opinion before the discovery of the Meissner effect was quite natural because of the analogy between the transition of a superconductor to the normal state and the very rapid increase of the magnetic field in the normal state, which Hirsch describes \cite{Hirsch2024}. In the second case, there is no doubt that the surplus work generates Joule heat, because only the energy of the magnetic field can be returned to the power source. W.H. Keesom and other physicists changed their opinion because the energy spent on the surplus work can be returned to the power source thanks to the Meissner effect. It is needed to know only Faraday's law in order to understand that the Meissner effect can be used to charge a battery: the amount of the negative work $W_{ns} = -\mu _{0}H_{2}^{2}$ performed at $H_{2} = H_{c}(T_{2})$ due to the Meissner effect equals twice the positive work $W_{aG1} = V\mu _{0}H_{2}^{2}/2 = VE_{m}$ performed during process 1 of the counterclockwise Gorter cycle. Thus, the negative surplus work $W_{surp-} = W_{aG1} + W_{ns} =  E_{m} - 2E_{m} = -V\mu _{0}H_{2}^{2}/2$ can increase the energy of the battery in each counterclockwise Gorter cycle. The increase in the energy $\Delta E_{bat} = N_{cyc}V\mu _{0}H_{2}^{2}/2$ is proportional the number of the cycles $N_{cyc}$, the volume of the superconductor $V$ and the square of magnetic field $H_{2}^{2}$. 
   
J. Hirsch agrees that the negative surplus work $W_{surp-} = -\mu _{0}H_{c}^{2}/2$ can be used to charge a battery. Moreover, he agrees that the energy of the battery will increase due to heat rather than free energy \cite{private2025}. But J. Hirsch is sure that the Gorter cycle is an analogy of the Carnot cycle: "{\it The process discussed here is completely analogous to a Carnot engine, where the system absorbs heat $Q_{1}$ from a heat reservoir at higher temperature, releases heat $Q_{2}$ to a heat reservoir at lower temperature, and delivers work $W = Q_{1} - Q_{2}$. The only difference is that in the Carnot cycle only two heat reservoirs at two different temperatures are needed, while in the Gorter cycle an infinite number of reservoirs at temperatures in the interval $T_{1} \leq T \leq T_{c}$ are needed}" \cite{Hirsch2024}. This analogy is not only false, but also testifies to the misunderstanding of the essence of the Carnot principle by those who use such an analogy.    

To avoid inevitable possibility of a perpetual motion machine, Carnot postulated the irreversibility of the process of converting energy into heat. This irreversibility is not absolute. A part $Q_{he} - Q_{co}$ of the heat $Q_{he}$ obtained in the heater can be converted in an energy $W$ if the other part $Q_{co} =  Q_{he} - W$ is transfered from the heater with a higher temperature $T_{he}$ to the cooler with $T_{co} < T_{he}$. Clausius formulated the Carnot principle as the second law of thermodynamics: the heat $Q_{W} = Q_{he} - Q_{co}$ having a non-zero entropy $S_{W} = Q_{W}/T_{he}$ can be converted to an energy $W$ if the decrease in entropy $\Delta S_{de} = - S_{W} =  -(Q_{he} - Q_{co})/T_{he} = -W/T_{he}$ is compensated by the entropy increase $\Delta S_{in} = Q_{co}/T_{co} - Q_{co}/T_{he} = (Q_{he} - W)/T_{co} - (Q_{he} - W)/T_{he}$ due to the transfer of the other part of the heat $Q_{co} =  Q_{he} - W$ from the heater with a higher temperature $T_{he}$ to the cooler with a lower temperature $T_{co}$. 

The falsity of Hirsch's analogy between the Gorter cycle and the Carnot cycle is no less obvious than the falsity of the claim of C.J. Gorter and H. Casimir \cite{Gorter1934}. This analogy is false already because in the Gorter cycle, energy is obtained from heat at the minimum temperature $T_{2} < T_{c}$, not at the maximum temperature $T_{he} > T_{co}$. No process of a heat transfer from the heater to the cooler is in the Gorter cycle, which could compensate for the entropy decrease on the amount $\Delta S_{de} = W_{surp-}/T_{2} = -\mu _{0}H_{c}^{2}/2T_{2}$. Hirsch himself writes about the absence of the irreversible process of the entropy increase in the Gorter cycle: "{\it We assume that all the heat exchanges with the environment happen under conditions where the system and the environment are at the same temperature, so that no net entropy is generated by the heat transfer process}" \cite{Hirsch2024}.  

J. Hirsch insists on his analogy despite its obvious falsity because of his illusion that he was able to prove, simply by writing his equation (1), what C.J. Gorter and H. Casimir obtained only on the base of "{\it the assumption, that the second law of thermodynamics applies also to the transition}" at $H_{2} = H_{c}(T_{2})$ \cite{Gorter1934}: "{\it neither the environment nor the sample changed their entropy}" \cite{Hirsch2024} in the Gorter cycle, as in the Carnot cycle. J. Hirsch did not realize that his proof was based on the assumption that he was proving. He did not realize also that his analogy between the Gorter cycle and the Carnot cycle contradicts his claim after equation (27b) that "{\it NO energy is dissipated as Joule heat}" \cite{Hirsch2024}. J. Hirsch contradicts himself since he ignores the contradictions between the books on superconductivity, which were provoked by the desire to save faith in the second law of thermodynamics. The analogy with the Carnot cycle is the only way to salvage this faith if the energy of the battery can increase due to heat in each counterclockwise Gorter cycle. But this is the most hopeless way.

\section{Conclusion}
Scientists must pursue the Truth, even if the truth is unpleasant. This article draws attention to unpleasant truth. The false thermodynamics of superconductors provoked by belief in thermodynamics shows that blind faith and even superstition can play an important and even decisive role in science. Einstein understood that thermodynamics is nothing more than a systematic answer to the question: what should be the laws of nature in order for a perpetual motion machine to be impossible. Even Einstein did not question the faith in the impossibility of a perpetual motion machine. Therefore, he was sure that "{\it Classical thermodynamics is the only physical theory of universal content, in respect which I am convinced that, within the framework of applicability of its basic concepts, it will never by overthrown}" \cite{EinsteinAutob}. 

The second law of thermodynamics has always been a matter of faith rather than understanding. This faith was the most blind in the 20th century. Only a few scientists understood that the impossibility of a perpetual motion machine required certain conditions, such as the assumption of molecular disorder \cite{Planck}. Therefore, almost everyone ignores the experimental evidences \cite{LP1962,Science2007,Letter2007,toKulik2010,APL2016,PCScien09,PCPRL09}  that violate this assumption. M. Smoluchowski was proving in 1914 \cite{Smoluchowski} that an useful perpetual motion machine cannot be possible because a Brownian particle cannot increase its potential energy by heat on a macroscopic amount much larger the energy of fluctuations $k_{B}T$. Many scientists have observed that a magnet with a macroscopic mass $M$ rises above a bulk superconductor to a height $h$ at its transition to the superconducting state. The potential energy of this magnet increases by a macroscopic amount $ghM \gg k_{B}T$ due to the Meissner effect. 

No one who has observed this levitation effect has asked the seemingly obvious question, "What energy should this macroscopic potential energy $ghM \gg k_{B}T$ be converted when the superconductor returns to the normal state and a magnet falls on its surface?" It should be emphasized that this energy should convert into heat not only according to the conventional theory of superconductivity \cite{BCS1957,GL1950}, which contradicts the second law of thermodynamics \cite{Hirsch2024,Entropy2022}. It is well known that when one body falls on the surface of another body, its potential energy, which has become kinetic energy, is converted into heat. If the macroscopic potential energy $ghM \gg k_{B}T$ is converted into heat when the superconductor returns to the normal state, then this energy must arise from heat during the transition to the superconducting state. Thus a perpetual motion machine, the impossibility of which M. Smoluchowski was proving \cite{Smoluchowski}, is observed in the levitation effect due to the Meissner effect. But most scientists are convinced of the impossibility of a perpetual motion machine, even when they observe it. 

Because of this unfounded belief, most results questioning the second law of thermodynamics are usually not published. The few publications \cite{book2005Capek,Fu1982,Vlada1999,AIP2002,Entropy2004,Peter2010,Daniel2023Lee,Lee2024Symmetry,Lee2024Energies} questioning the common belief are ignored by most scholars. It is important to draw reader's attention that blind faith in thermodynamics has provoked a false thermodynamics of superconductors and a regression in the understanding of thermodynamics by superconductivity experts. Only because of the blind faith W.H. Keesom could forget that no work can be done at a phase transition when he derived the equations (15) for the latent heat in 1924. Only because of the blind faith W.H. Keesom and others failed to understand that the negative surplus work performed in each Gorter cycle due to the Meissner effect violates the second law of thermodynamics, since free energy cannot change in a closed cycle. 

The blind faith provoked the false thermodynamics of superconductors developed by Gorter and Casimir and three logically possible contradictions with the basics of thermodynamics in books on superconductivity. Physicists must finally admit that they cannot avoid violation of the second law of thermodynamics by the Meissner effect without contradicting the law of conservation of energy. Superconductivity experts should understand why the thermodynamics of Gorter and Casimir \cite{Gorter1934} is false and the reason for the contradiction between the books on superconductivity if only to avoid repeating the mistakes made by J. Hirsch \cite{Hirsch2024}. Physicists should know that the false thermodynamics of superconductors was provoked by the blind faith in the second law of thermodynamics. The problem of the perpetual motion machine as a heat engine in which heat is converted into energy without any fuel should lastly cease to be the object of faith and become the object of understanding for most scientists. Getting scientists rid of the centuries-old prejudice about the impossibility of a perpetual motion machine may have not only fundamental but also great practical importance. In conclusion, I would like to draw the attention of experimenters that they can corroborate experimentally the possibility to increase the energy of a battery by the amount $\Delta E_{bat} = N_{cyc}V\mu _{0}H_{2}^{2}/2$ with $N_{cyc}$ counterclockwise Gorter cycles.      

This work was made in the framework of State Task No 075-00295-25-00.

\end{document}